\documentclass[fleqn,usenatbib]{mnras}

\usepackage{newtxtext,newtxmath}

\usepackage[T1]{fontenc}
\usepackage{ae,aecompl}

\usepackage{graphicx}	
\usepackage{amsmath}	
\usepackage{amssymb}	
\usepackage{xcolor}

\newcommand{\rsun}{R_{\odot}} 
\newcommand{\mjup}{M_{\text{Jup}}} 
\newcommand{\rjup}{R_{\text{Jup}}} 

\title[Informed priors with ML classification]{Optimizing exoplanet atmosphere retrieval using unsupervised machine-learning classification}

\author[J.J.C. Hayes et al.]{
J.J.C. Hayes,$^{1}$\thanks{E-mail: joshjchayes@gmail.com}
E. Kerins,$^{1}$
S. Awiphan,$^{2}$
I. McDonald,$^{1}$
J.S. Morgan,$^{1}$
\newauthor
P. Chuanraksasat,$^{2}$ 
S. Komonjinda,$^{3,4}$
N. Sanguansak, $^{5}$
and P. Kittara$^{6}$ (SPEARNET)
\\
$^{1}$Jodrell Bank Centre for Astrophysics, University of Manchester, Oxford Road, Manchester, M13 9PL, UK\\
$^{2}$National Astronomical Research Institute of Thailand, 260 Moo 4, Donkaew, Mae Rim, Chiang Mai, 50180, Thailand\\
$^{3}$Research Center in Physics and Astronomy, Faculty of Science, Chiang Mai University, Chiang Mai, 50200, Thailand\\
$^{4}$Department of Physics and Materials Science, Faculty of Science, Chiang Mai University, Chiang Mai, 50200, Thailand\\
$^{5}$School of Physics, Institute of Science, Suranaree University of Technology, 111 University Ave., Suranaree, Nakhon Ratchasima, 30000, Thailand\\
$^{6}$Department of Physics, Faculty of Science, Mahidol University, Bangkok 10400, Thailand
}

\date{Accepted XXX. Received YYY; in original form ZZZ}

\pubyear{2019}

\begin{document}
\label{firstpage}
\pagerange{\pageref{firstpage}--\pageref{lastpage}}
\maketitle

\begin{abstract}
One of the principal bottlenecks to atmosphere characterisation in the era of all-sky surveys is the availability of fast, autonomous and robust atmospheric retrieval methods. We present a new approach using unsupervised machine learning to generate informed priors for retrieval of exoplanetary atmosphere parameters from transmission spectra. We use principal component analysis (PCA) to efficiently compress the information content of a library of transmission spectra forward models generated using the PLATON package. We then apply a $k$-means clustering algorithm in PCA space to segregate the library into discrete classes. We show that our classifier is almost always able to instantaneously place a previously unseen spectrum into the correct class, for low-to-moderate spectral resolutions, $R$, in the range $R~=~30-300$ and noise levels up to $10$~per~cent of the peak-to-trough spectrum amplitude. The distribution of physical parameters for all members of the class therefore provides an informed prior for standard retrieval methods such as nested sampling. We benchmark our informed-prior approach against a standard uniform-prior nested sampler, finding that our approach is up to a factor two faster, with negligible reduction in accuracy.  We demonstrate the application of this method to existing and near-future observatories, and show that it is suitable for real-world application. Our general approach is not specific to transmission spectroscopy and should be more widely applicable to cases that involve repetitive fitting of trusted high-dimensional models to large data catalogues, including beyond exoplanetary science.
\end{abstract}




\section{Introduction}
\label{sec:intro}
Transmission spectroscopy is a significant and growing branch of exoplanetary research. Since the first detections and studies of the atmosphere of HD~209458~b \citep{charbonneau2002detectionHD209458b_atmosphere, vidal2003HD209458b_atmosphere}, transmission spectroscopy has become the leading diagnostic tool for the exploration of the chemistry and physics of exoplanetary atmospheres. The technique has been used to detect a variety of chemical species, including potassium \citep{potassium2011, potassium2015}, sodium \citep{redfield2008sodium_detection}, water \citep{tinetti2007water, konopacky2013CO_H2O_detection, Birkby_water_2013}, titanium oxide \citep{sedaghati2017TiOdetection, NugruhoTiO}, carbon monoxide \citep{Snellen_CO_2010, Brogi_CO_2012}, HCN \citep{Hawker_HCN}, methane \citep{guilluy_CH4}, helium \citep{Allart_He}, titanium, and iron \citep{Hoeijmakers_Ti_Fe}.

Currently there are almost 3000 confirmed planets detected through primary transit\footnote{The Extrasolar Planet Encyclopedia: \url{http://exoplanet.eu}}. The majority of these were discovered by \emph{Kepler} \citep{kepler}, though most of the hosts are too faint for current atmospheric follow-up.

With the launch of the \emph{Transiting Exoplanet Survey Satellite} \citep[\emph{TESS},][]{TESS_paper} and the development of ground-based transit surveys like the \emph{Next Generation Transit Survey} \citep[NGTS,][]{NGTS}, both of which target much brighter hosts than \emph{Kepler}, there will be an increase of several orders of  magnitude in the number of potential transmission spectroscopy targets that can be followed-up by both ground- and space-based observatories. 

With such a step change in the number of transmission spectroscopy targets it will become necessary both to prioritise targets for transmission spectroscopy studies and to implement autonomous and efficient methods for modelling their spectra.

The \emph{Spectroscopy and Photometry of Exoplanetary Atmospheres Research Network} ({SPEARNET}), is developing a network and tools to respond to the paradigm shift from a target-starved to a asset-starved  era. \cite{morgan2019metric} presents a Decision Metric that can be used to optimally pair transmission spectroscopy targets to telescopes within a heterogeneous globally-distributed network. One of the primary science drivers for the use of a Decision Metric is that the selected targets can be used for statistically robust population studies as it removes unquantified bias in manual expert selection.

However, even with an optimised selection approach, the expected increase in the number of transmission spectroscopy observations requires improvements in the efficiency of atmospheric retrieval methods, which are typically computationally expensive and time consuming. A variety of models are now available, given the recent push towards producing open source forward models and retrieval codes, using various different methods. ATMO \citep{ATMOGoyal2018} uses a pre-computed scalable grid to calculate transmission spectra, but does not provide a built-in retrieval module. Tau-REx \citep{Tau-REX_I, Tau-REx_retrieval} is an open source package which provides a Bayesian retrieval framework for transmission spectra, complete with a highly detailed forward model. petiRADTRANS \citep{petitRADTRANS} is a forward model for both transmission and emission spectra at moderate ($R = 1000$) and high ($R>10^6$) spectral resolutions, but does not come packaged with an inbuilt retrieval module. Exo-Transmit \citep{exotransmit} is a forward model, which has recently been streamlined and packaged with a retrieval module within the PLATON \citep{PLATON2019} Python package. 

In this paper, we use unsupervised machine learning to aid and improve the efficiency of exoplanetary atmosphere model retrieval. In the era of ``big data'' the problem of mining and characterising large datasets is not unique to exoplanetary science, nor even to astrophysics. We believe that our general method should find broad applicability to cases that involve repetitive fitting of trusted high-dimensional models to large data catalogues.

We give a brief overview of nested sampling methods and discuss the appeal of informed priors in Section \ref{sec:nested_sampling}. We introduce the machine learning algorithm used to generate informed priors and demonstrate how to apply it to a set of simulated transmission spectra in Section \ref{sec:classification}. In Section~\ref{sec:Use_in_retrieval} we discuss the effects of applying optimised priors to nested sampler retrieval of exoplanetary atmosphere parameters. We offer some conclusions on our approach in Section \ref{sec:conclusions}.

\section{Nested sampling and priors}
\label{sec:nested_sampling}
In this paper, we primarily focus on using nested sampling algorithms, implemented using the Python package \texttt{dynesty} \citep{speagle2018dynesty}. Nested sampling has been shown to be an effective retrieval method when dealing with models of high dimensionality. In order to convey the advantage of using informed priors for model retrieval, we present here a brief overview of the nested sampling method. This is not exhaustive, and we direct the interested reader to the work of \cite{nested_sampling_original2004}, or the documentation provided by \texttt{dynesty}\footnote{\url{https://dynesty.readthedocs.io/en/latest/overview.html}} for further details.

As in most Bayesian modelling, the primary aim of nested sampling is to estimate the posterior likelihood $P\left(\mathbf{\Theta} | \mathbf{D}, M \right)$ of a set of parameters $\mathbf{\Theta}$ for a model $M$ given a set of data $\mathbf{D}$. We can then use Bayes' rule to give 
\begin{equation}
    P\left(\mathbf{\Theta} | \mathbf{D}, M \right) = \frac{P\left(\mathbf{D} | \mathbf{\Theta}, M \right)~P\left(\mathbf{\Theta} | M \right)}{P\left(\mathbf{D} | M \right)} \equiv \frac{\mathcal{L}\left(\mathbf{\Theta}\right)~\pi\left(\mathbf{\Theta}\right)}{\mathcal{Z}}.
\end{equation}
Here, $\mathcal{L}\left(\mathbf{\Theta}\right)$ is the likelihood and $\pi\left(\mathbf{\Theta}\right)$ is the prior. $\mathcal{Z}$ is known as the evidence, and can be written as 
\begin{equation}
    \mathcal{Z} = \int_{\mathbf{\Theta}} \mathcal{L}\left(\mathbf{\Theta}\right)~\pi\left(\mathbf{\Theta}\right) d\mathbf{\Theta}
    \label{eq:evidence}
\end{equation}
where the integral is taken over the range of $\mathbf{\Theta}$.

Many Bayesian inference methods focus on sampling the posterior and leave the evidence as a by-product. Nested sampling reverses this and concentrates on sampling the evidence, leaving the posterior sampling as a secondary focus. 

Generally, $\mathcal{Z}$ cannot be directly evaluated with computational efficiency. Instead, in nested sampling the evidence is estimated by integrating the prior in nested shells of constant likelihood. Rather than integrating over all $\mathbf{\Theta}$, nested sampling estimates $\mathcal{Z}$ by approaching the integral in Equation \ref{eq:evidence} as an integral over the prior volume
\begin{equation}
    X\left(L\right) \equiv \int_{\mathbf{\Theta}\colon\mathcal{L}\left(\mathbf{\Theta}\right) > L} \pi\left(\mathbf{\Theta}\right) d\mathbf{\Theta}
\end{equation}
which is contained within an iso-likelihood contour set $\mathcal{L}\left(\mathbf{\Theta}\right) = L$. $X\left(L\right)$ can be thought of as the amount of prior mass associated with likelihoods greater than $L$. It is defined such that 
$X\left(0\right)~=~1$, $X\left(L_\text{max}\right)=0$, and $dX$ is the prior mass associated with the likelihoods between $L$ and $L + dL$. This can then be used to calculate the evidence through the relation
\begin{equation}
    \mathcal{Z} = \int^{\infty}_{0} X \left(L\right) dL = \int^1_0 \mathcal{L}\left(X\right) dX
    \label{eq:shell_integrals}
\end{equation}
where $\mathcal{L}\left(X\right)$ is the inverse of $X\left(L\right)$. Through evaluating the iso-likelihood contours with a set of $X_i$ values
\begin{equation}
    1 = X_0 > X_1 > ... > X_N > 0, 
\end{equation}
the evidence can be computed by solving Equation \ref{eq:shell_integrals} using standard numerical methods. 

The basic algorithm of nested sampling is as follows:
\begin{enumerate}
    \item Draw a set of $K$ ``live points'' from the prior. 
    \item For every iteration $i$, find the likelihood $L_i$ of each live point and remove the point with the lowest likelihood, which now becomes a ``dead point.''
    \item Replace the dead point with a new live point sampled from the prior, but with the condition that $L_{i+1} > L_i$. 
\end{enumerate}
\cite{nested_sampling_original2004} shows that the average remaining prior mass $X_i$ is given by 
    \begin{equation}
        \log X_i = -\frac{i \pm \sqrt{i}}{K}.
        \label{eq:prior_volume}
    \end{equation}
We can therefore estimate the evidence integral in Equation \ref{eq:shell_integrals} using a set of $N_\textrm{dead}$ dead points through 
\begin{equation}
    \mathcal{Z} \approx \hat{\mathcal{Z}} = \sum^{N_\textrm{dead}}_{i=1} f\left(L_i\right) g\left(\Delta X_i\right).
\end{equation}
The functions $f\left(L_i\right)$ and $g\left(\Delta X_i\right)$ will vary depending on the integration scheme being used. In the case of \texttt{dynesty}, which uses a trapezium rule integration, $f(L_i) = (L_{i-1} + L_i) / 2$ and $g\left(\Delta X_i\right) = X_{i-1} - X_i$.

In theory, this algorithm could iterate forever, sampling a smaller and smaller likelihood volume. Consequently, a stopping criterion must be introduced. A straightforward criterion is to continue until there is only a certain fraction of evidence left to be sampled. At a given iteration $i$, by assuming that the remaining prior mass within the shell containing the last dead point is a uniform likelihood region, the remaining evidence can be approximated by 
\begin{equation}
    \Delta\hat{\mathcal{Z}}_i \approx L_\textrm{max} X_i.
\end{equation}
A stopping criterion can then by defined by using the log-ratio between the current estimated evidence $\hat{\mathcal{Z}}_i$ and $\Delta\hat{\mathcal{Z}}_i$, given by 
\begin{equation}
    \Delta \ln \hat{\mathcal{Z}}_i \equiv \ln \left(\hat{\mathcal{Z}}_i + \Delta \hat{\mathcal{Z}}_i \right) - \ln \hat{\mathcal{Z}}_i.
    \label{eq:stopping_criterion}
\end{equation}
This gives an estimate for the remaining fraction of the evidence that has not been sampled. A basic stopping criterion of a nested sampler is to sample until only a given fraction is left. \texttt{dynesty} uses the stopping criterion $\Delta\mathcal{\hat{Z}} = (K-1)\times10^{-3} + 10^{-2}$, where $K$ is the number of live points drawn from the prior. Once the stopping criterion is met, the \texttt{dynesty} algorithm takes all the live points and redistributes them uniformly across the surviving posterior space to further increase the evidence sampling. This tends to result in $>99.9$~per~cent of the total evidence being sampled.

We can see from Equation \ref{eq:evidence} that the priors are involved in the calculation of $\mathcal{Z}$, and the nature of the priors has a significant impact upon the behaviour of a nested sampling routine, most notably in the run time. In general, the posterior volume $P\left(\mathbf{\Theta}\right)$ is more localised than the prior volume $\pi\left(\mathbf{\Theta}\right)$. By using the Kullback-Leiber divergence, $H$, \citep{Kullback59}, the amount of information gained from updating the prior to the posterior is given by 
\begin{equation}
    H \equiv \int_\mathbf{\Theta} P\left(\mathbf{\Theta}\right) \ln \left(\frac{P\left(\mathbf{\Theta}\right)}{\pi\left(\mathbf{\Theta}\right)}\right) d\mathbf{\Theta}.
    \label{eq:KL-diverence}
\end{equation}
\cite{nested_sampling_original2004} shows that the number of iterations, $N_\textrm{iter}$, required for the updated prior to be equivalent to the posterior goes as
\begin{equation}
    N_\textrm{iter} \propto KH.
    \label{eq:iterations-prior_relation}
\end{equation}
Since a narrower prior leads to a smaller value of $H$, a more informed prior is desirable since fewer iterations are required to obtain a result. Whilst this relationship shows that a low number of live points, $K$, is desirable, \cite{nested_sampling_original2004} notes that in practice this leads to a very fast sampling which skips most of the bulk of the posterior volume. \cite{nested_sampling_original2004} recommends that a larger number of live points is preferable and safer to use, as this leads to gentler progress and more complete sampling of the posterior.

\section{Informed priors from unsupervised machine learning}
\label{sec:classification}

\subsection{Why unsupervised machine learning?}
\label{sec:why?}
In modern astronomy, we frequently encounter situations where a trusted multi-parameter model is applied repeatedly to large data catalogues in order to calculate model parameters for each object in the catalogue. In such cases it can be more economical to compute a set of informed priors that can be used to make such bulk model retrievals more efficient. 

Equations \ref{eq:KL-diverence} and \ref{eq:iterations-prior_relation} show that, in the case of nested sampling retrieval, narrow, more informed priors are preferable to wide, uninformed priors. Priors for specific objects can often be informed by additional measurements that may be available for that object. But even where there is no such additional information available the variability of the trusted model itself across the parameter range may exhibit a natural taxonomy of behaviour. This opens the possibility of using a classification approach where a dataset is initially identified as belonging to a certain class of models. The distribution of parameter values within that class is then employed as an informed prior for retrieval of the best fit solution. 

This approach is preferable if three requirements are met: 
\begin{enumerate}
    \item the time taken to retrieve parameters for all members of the data catalogue is large compared to the time taken to initially segregate the model into classes;
    \item members of the data catalogue can be classified correctly with high reliability;
    \item the time taken to classify a member of the data catalogue is significantly outweighed by the subsequent time saved by parameter retrieval through an informed-prior approach compared to  the standard uninformed (uniform) prior approach.
\end{enumerate}
In Section~\ref{sec:Use_in_retrieval} we show that our approach allows all of these requirements to be met for rapid atmospheric forward modelling codes like PLATON.

Unsupervised machine learning (ML) classification can be used to classify models and to retrieve a set of priors for a given class. For transmission spectroscopy retrieval we find that, once trained, our classifier is able to identify which class a set of data belongs to and generate priors for that class in generally less than a second. By using unsupervised machine learning we allow the model behaviour itself to determine the taxonomy of the classification though, as discussed in Section \ref{sec:informed_priors_results}, some care is required in determining the allowed number of classes. 

In this section, we introduce our algorithm for generating informed priors in the context of transmission spectroscopy of exoplanetary atmospheres. This is an area which has already seen some application of machine learning, for example through the use of neural networks \citep{Cobb12019NeuralNetEnsemble} or supervised machine learning algorithms \citep{marquez2018supervised}. These methods generally look to the machine learning algorithm to do the full retrieval of atmospheric parameters from data. We suggest that by using machine learning to inform more traditional retrieval techniques, rather than relying on the machine learning algorithm to do the entire data processing, one is in a ``best of both worlds'' situation, where we combine the pattern recognition and statistical analysis abilities of machine learning with the ability to more finely sample parameter space of traditional Monte Carlo retrieval techniques. Through this method, coupled with a sensible choice of number of classes based on the quality of the data, we are able to minimise the risk of misclassification leading to incorrect results, which is a major issue in retrieval which relies solely upon ML classification.

\subsection{Simulating transmission spectra for classification}
\label{sec:spectra_simulation}
Our algorithm consists of using an unsupervised ML classifier to identify classes of transmission spectra and generate parameter distributions for each class. We then use it to classify other spectra and recover the parameter distributions of their class for use as our informed priors for parameter retrieval.

In order to train the classifier, we require a set of training data with known parameters. As we are using an unsupervised algorithm to give the classifier as much freedom to define classes as possible, we do not pass the known parameters to the classifier, but we do track them so that we can build the prior distributions for each resulting class. The performance of our algorithm is strongly linked to the accuracy of the underlying model which is used to simulate the training data. However, using PLATON we find that the total time taken to simulate all of our training data and to train the classifier with it generally takes less time than running a single model retrieval with uninformed priors, easily fulfilling requirement (i) of Section~\ref{sec:why?} . We stress that a trained classifier can only operate on data with similar appearance to the data with which it was trained. Therefore it is necessary to train the classifier with a noiseless training set that has the same wavelength bins as the data we wish to classify. 

The training data which we use are simulated transmission spectra generated using \textsc{PLATON} \citep{PLATON2019}. We use a restricted subset of five variables for our tests: planet mass, atmospheric temperature, metallicity, carbon-oxygen ratio, and Rayleigh scattering multiplier. The Rayleigh scattering multiplier is defined such that 1 is Earth-like, and higher values imply proportionately larger scattering. We uniformly random sample the parameter space defined on the ranges in Table \ref{tab:parameter values}. The other variables that can be adjusted within PLATON (host star radius, planet radius and cloud pressure) are all fixed. The host star and planet radii are fixed at $1~\rsun$ and $1~\rjup$, respectively, and we assume a cloud-free atmosphere. 

We fix the planet and host radius as these do not substantially affect the shape of the spectrum, only its amplitude, and, in any case, they are usually already accurately known for a planet that is being targeted for transmission spectroscopy. We are therefore concerned here with training the classifier to recognise changes in spectral morphology, rather than overall amplitude. We assume a cloud-free atmosphere for much the same reason. The effect of the presence of substantial high altitude cloud is to mask spectral features, making morphological classification more difficult. In this case we should expect the set of informed priors provided by our approach to simply tend towards those of uninformed priors, i.e. all spectra will appear to the classifier to belong to one class that spans the entire parameter range. The outcome will be that the model retrieval efficiency will be reduced to that of the standard approach. In the case where morphology information is available from a spectrum, however, we seek to develop an approach that can use this information to classify the spectrum before further analysis and therefore potentially provide more efficient retrieval.

For each set of parameters sampled, we use PLATON to generate the associated spectrum. The spectrum is then convolved to a given spectral resolution, $R$, keeping the full wavelength range of PLATON, namely logarithmically spaced values in the range $0.3~\umu$m to $30~\umu$m. For this investigation we use $R=30,~100$, and $300$ to represent low-, and medium-resolution transmission spectroscopy. There is no technical limitation in the method preventing the use of even higher spectral resolutions up to the limit of PLATON itself, $R=1000$. Similarly, one could use a combination of resolutions across different wavebands, or a restricted range of wavelengths, which we discuss further in Section~\ref{sec:real_spectra}.  Also, our method does not specifically require the use of PLATON to simulate the spectra. Any other model that is fast enough to satisfy the requirements of Section~\ref{sec:why?} can be used, provided that the model is characterised by continuous parameters that allow the user to generate an appropriate set of training data.

\subsection{Generating informed priors}
\label{sec:training_classifier}

To train the classifier, we sample $100\,000$ points in the five-dimensional parameter space of planet mass, atmospheric temperature, metallicity, carbon-oxygen (C/O) ratio, and scattering factor. Each sample in parameter space is then used as a parameter set and an associated transmission spectrum is simulated using PLATON, which is then binned to give the desired spectral resolution. 

For unsupervised ML classification the classifier does not know anything about the underlying parameters that control the morphology of the spectrum, and is even unaware of how many parameters there are. Each spectrum is an array of flux values and the classifier treats each element of the array as a realisation from a \emph{pseudo-dimension} that spans a continuous space of possible values. 

\begin{figure}
    \centering
    \includegraphics[width=\columnwidth]{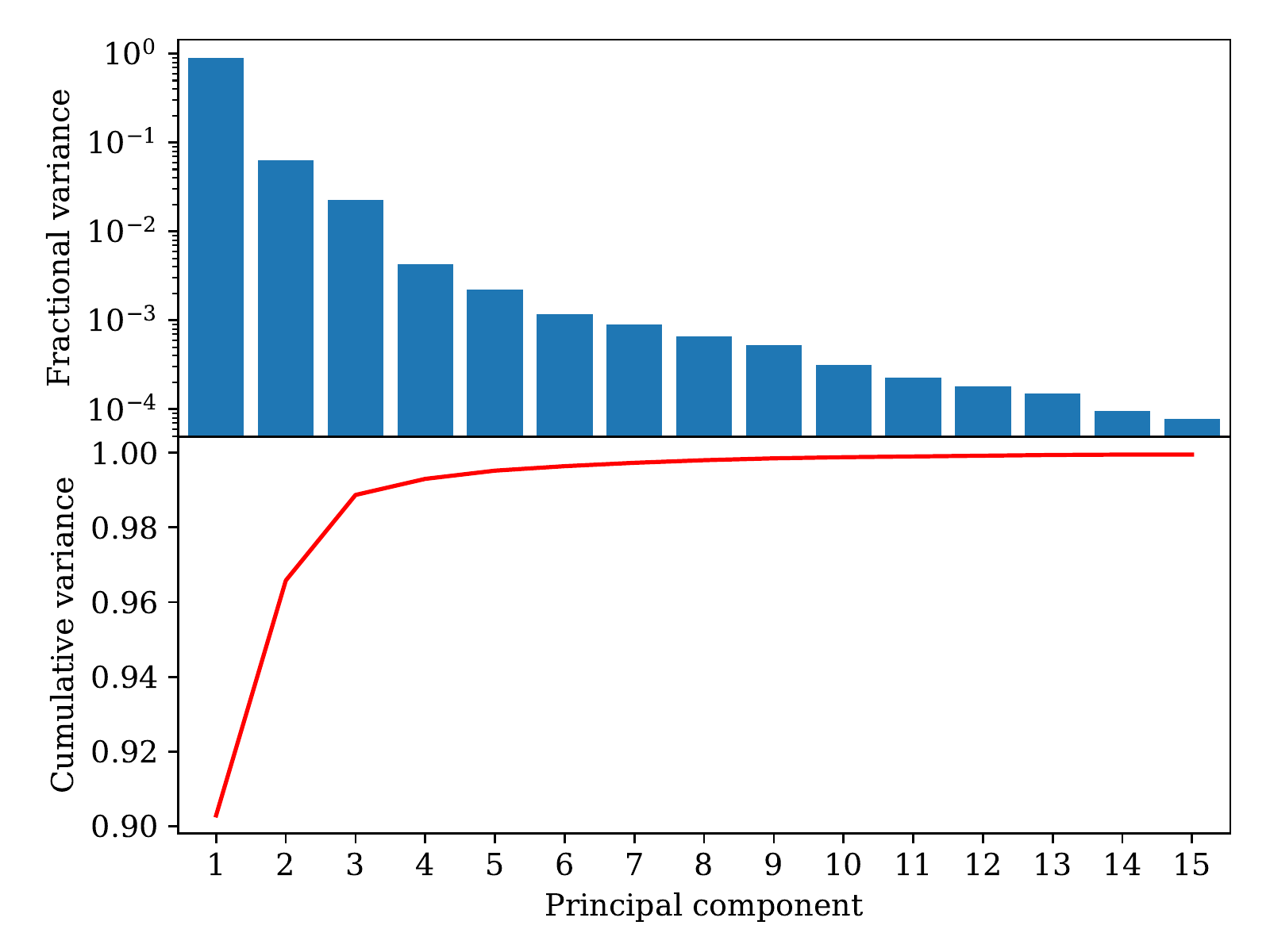}
    \caption{\textit{Top:} The fraction of total variance provided by each of the first 15 principal components. This PCA was conducted on 100,000 spectra at a spectral resolution $R=100$, sampled uniformly from the parameter values detailed in Table \ref{tab:parameter values}. \textit{Bottom:} The cumulative variance of the principal components. $99.9$~per~cent of the variance of the spectra is contained within the first ten components.}
    \label{fig:pca_variance}
\end{figure}

Each resulting spectrum can therefore be thought of as a single coordinate in a spectral space with $N_\textrm{dims}$ pseudo-dimensions, where the spectrum $\mathbf{s}$, which is produced by a set of parameters $\mathbf{p}$, is given by 
\begin{equation}
    \mathbf{s}\left(\mathbf{p}\right) = \left[\Delta F(\lambda_1),\Delta F(\lambda_2),...,  \Delta F(\lambda_{N_\textrm{dims}})\right]
\end{equation}
where $\Delta F\left(\lambda_i\right)$ is the transit depth at wavelength $\lambda_i$. The pseudo-dimensionality of the spectral space is determined by the number of wavelength bins, $N_\textrm{bins}$, required to obtain a given spectral resolution $R$:
\begin{equation}
    N_\textrm{{dims}} = N_\textrm{bins} = R \ln \left(\frac{\lambda_\text{max}}{\lambda_\text{min}}\right).
    \label{eq:dimensionality}
\end{equation}
Binning a full PLATON spectrum to $R=100$ results in a spectral space of 461 pseudo-dimensions. We use tools provided by scikit-learn \citep{scikit-learn} to find clusters of similar spectra. However, as discussed in the scikit-learn documentation\footnote{\url{https://scikit-learn.org/stable/user_guide.html}}, clustering algorithms are very slow for high dimensionality data. In order to reduce the dimensionality, we conduct a principal component analysis (PCA). Figure \ref{fig:pca_variance} shows the fraction of the total variance contained within each of the first 15 principal components for $R=100$, and shows that $99.9$ per cent of the variance is contained within the first ten components. Keeping these 10 components allows us to reduce the pseudo-dimensionality by a factor of almost 50, with negligible information loss. 

Once the PCA is completed, an unsupervised $k$-means clustering algorithm is run to find clusters in the reduced-dimensionality PCA space. These clusters are used to define a set of spectral classes that any new spectrum can be compared to and classified by. For $k$-means clustering we must select the allowed number of classes. We use 30 classes, though we note that there can be variation in the results of our method depending upon both the allowed number of classes and the number of spectral principal components which are used. We discuss the effects of the number of classes more in Section \ref{sec:classification_accuracy}. Whilst there are unsupervised ML classification schemes that do not require the user to pre-determine the number of classes, the simplicity of the $k$-means clustering method means that it is straightforward and fast to determine which class a new spectrum belongs to. This is a significant benefit when dealing with a potentially large catalogue of objects.

\begin{figure*}
    \centering
    \includegraphics[width=\textwidth]{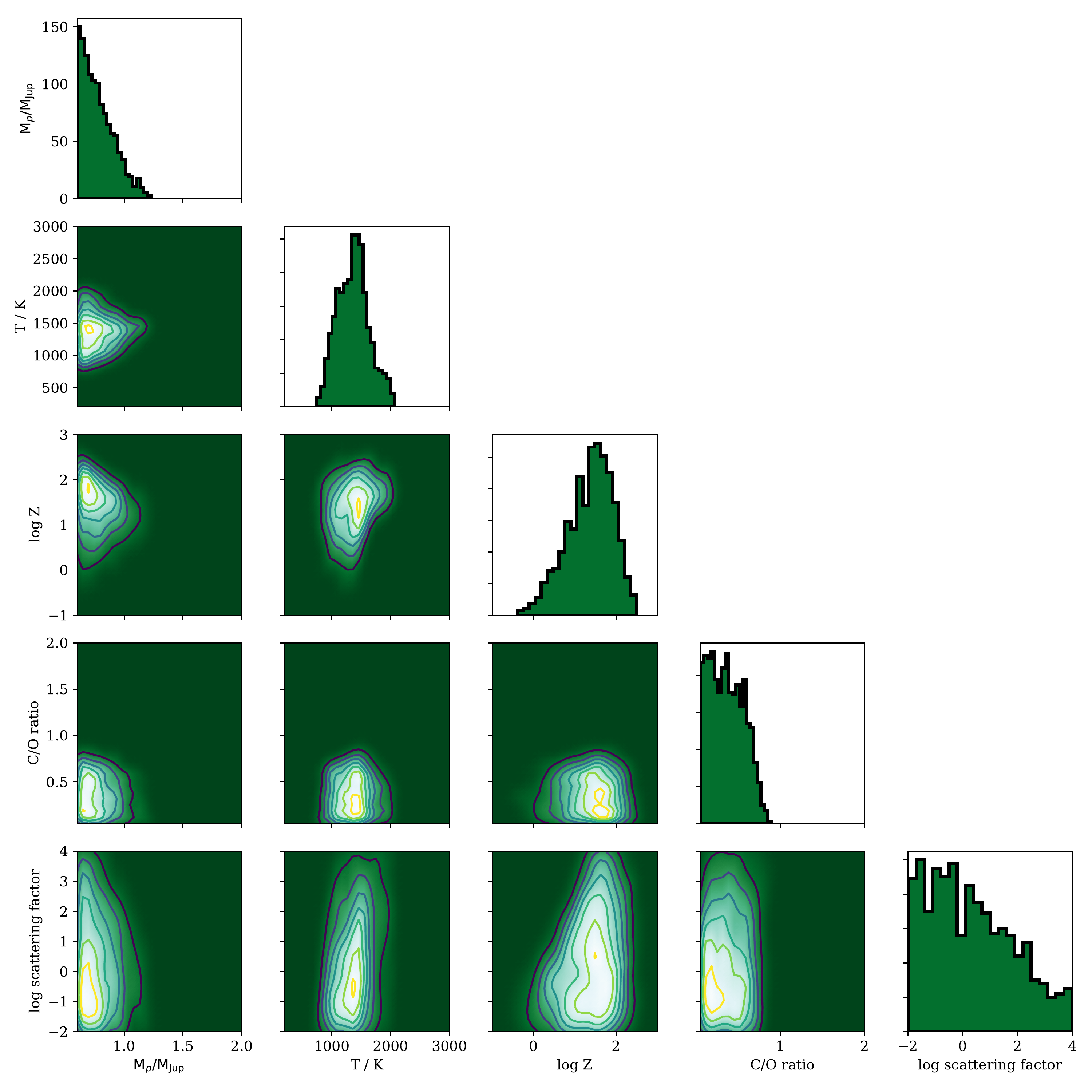}
    \caption{An example parameter distribution obtained from our clustering algorithm. The axis ranges are the full ranges stated in Table \ref{tab:parameter values}. The marginalised distributions can be used as informed priors for retrieval algorithms.}
    \label{fig:cluster_param_distribution1}
\end{figure*}

After the classes are determined in PCA space, the physical parameters of each member of a class are extracted. This allows the recovery of the distribution of class membership in physical parameter space, an example of which is shown in Figure~\ref{fig:cluster_param_distribution1}. It is clear from Figure~\ref{fig:cluster_param_distribution1} that the parameter distributions recovered from unsupervised ML classification occupy significantly concentrated volumes within the overall parameter space, as would be expected for spectra that exhibit similar morphology.

At this point a new spectrum with unknown parameters, $\mathbf{s}\left(\hat{\mathbf{p}}\right)$, can now be passed to the classifier and assigned to a class by converting $\mathbf{s}\left(\hat{\mathbf{p}}\right)$ into PCA space coordinates and finding the closest class centre. The parameter distribution of this class can then be used as an informed prior to aid the retrieval of $\hat{\mathbf{p}}$.

\begin{table}
    \caption{Parameter values over which the simulated spectra are generated using PLATON. Parameters marked as variable are uniformly sampled between the low and high values stated. Parameters which are not varied within this investigation have the relevant fixed value provided in the low column.}
    \label{tab:parameter values}
    \begin{tabular}{l|c|c|l}
        \hline
        Parameter & Variable? & Low & High \\
        \hline
        Star radius $\left(\rsun\right)$ & No & $1$ & - \\
        Planet mass $\left(\mjup\right)$& Yes & $0.6$ &$2$ \\
        Planet radius $\left(\rjup\right)$ & No & $1$ & - \\
        Atmosphere temperature (K) & Yes &$200$ & $3000$\\
        Log metallicity & Yes & $-1$ &$3$\\
        C/O ratio & Yes & $0.05$ & $2$\\
        Log cloudtop pressure ($\log$ Pa)& No & $\infty$ & -\\
        Log scattering factor & Yes & $-2$ & $4$\\
        \hline
    \end{tabular}

\end{table}

\section{Use of classifications in aiding retrieval}
\label{sec:Use_in_retrieval}

\subsection{Quantifying retrieval quality}
\label{sec:retrieval_accuracy}
We use the marginalised physical parameter distributions for each of the classes identified by the classifier as informed priors for the nested sampler routine in PLATON. We discuss the implementation and effect of this further in Section \ref{sec:informed_priors_results}. Before we discuss the effects, however, we must first define some metrics by which we can quantify the quality and efficiency of retrieval. 
In any retrieval, there are three properties which one seeks to optimise: accuracy, precision, and speed. 

To quantify speed, we use $N_\textrm{iter}$, the number of iterations required before the stopping criterion of the nested sampler is reached. The nested sampler retrieval in PLATON uses \texttt{dynesty} \citep{speagle2018dynesty} (as of v4.0). The stopping criterion used by PLATON is the default \texttt{dynesty} value of $\Delta\mathcal{\hat{Z}} = (K-1)\times10^{-3} + 10^{-2}$, which is the default when the final set of live points is included in the final calculation of the evidence and posteriors (c.f. Section~\ref{sec:nested_sampling}). We use the number of iterations rather than the actual computation time as a metric because this is machine agnostic; we are showing the potential of this technique rather than presenting concrete values on run times on a specific hardware setup. 

Calculation of accuracy and precision metrics relies on being able to find a distance between sets of parameters $\textbf{p}$. In order to calculate these we scale each parameter $p$ to a dimensionless value $p'$ between 0 and 1 using 
\begin{equation}
    p' = \frac{p - p_\text{min}}{p_\text{max}-p_\text{min}},
\end{equation}
where $p_\text{min}$ and $p_\text{max}$ are taken from Table \ref{tab:parameter values}. From here on we use prime notation to denote any coordinate or value which is unit normalised.

Using simulated data to test our classifier, for every model retrieval we have two sets of coordinates in parameter space: the true (simulated) set of parameters, $\mathbf{p}_t$, and the set of parameters obtained from the retrieval, $\mathbf{p}_r$. For our accuracy metric, $\mathcal{M}_1$, we use the Euclidean distance between these points in the unit-normalised physical parameter space:
\begin{equation}
    \mathcal{M}_1 = \lVert \mathbf{p}'_t - \mathbf{p}'_r \rVert.
    \label{eq:M1}
\end{equation}

PLATON retrieval returns an upper and lower $1\sigma$ uncertainty on each fitted parameter. For our five-parameter grid these error bars define a 5-D hyper-ellipsoid around $\mathbf{p}'_r$. Allowing for asymmetric errors by restricting consideration to one orthant of the hyper-ellipsoid, the surface of the $n$-dimensional hyper-ellipsoid bounds a set of coordinate points $x'_i$ such that
\begin{equation}
    \sum_{i=1}^{n} \left(\frac{x'_i - p'_{ri}}{\sigma'_i}\right)^2 \leq 1
\end{equation}
where the sum is taken over each coordinate $i$, $p'_{ri}$ is the retrieved value of that coordinate and $\sigma'_i$ is the associated uncertainty on $p'_{ri}$ in the considered orthant.

\begin{figure}
    \centering
    \includegraphics[width=\columnwidth]{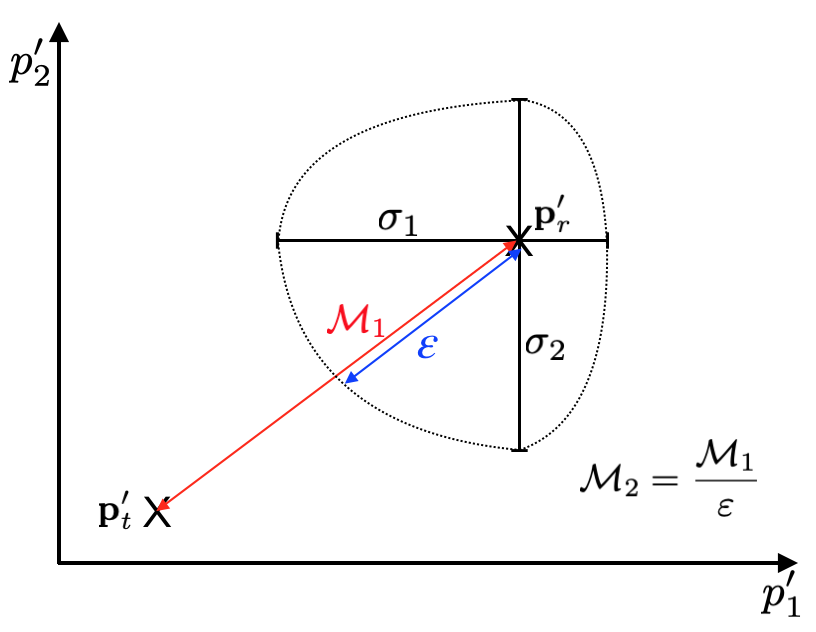}
    \caption{An illustration of the metrics used to quantify the quality of the results of a nested sampler retrieval in a dimensionless two-dimensional parameter space. $\mathcal{M}_1$ is an accuracy metric defined by the distance between the true set of parameters, $\mathbf{p}'_t$, and the retrieved set of parameters, $\mathbf{p}'_r$. The dashed line represents the composite error hyper-ellipsoid associated with $\mathbf{p}'_r$, and $\sigma_1$ and $\sigma_2$ are the retrieval errors in the  orthant considered in the calculation of $\varepsilon$, the $1\sigma$ distance in the direction of $\mathcal{M}_1$. $\mathcal{M}_2$ is defined as the ratio of $\mathcal{M}_1$ and $\varepsilon$.}
    \label{fig:metrics_illustration}
\end{figure}

The intercept, $\mathbf{x}'$, of the surface boundary of the relevant hyper-ellipsoid and the line joining $\mathbf{p}'_r$ and $\mathbf{p}'_t$ can be found by solving the set of simultaneous equations
\begin{equation}
    \mathbf{x}' -\mathbf{p}'_t - A\left(\mathbf{p}'_r - \mathbf{p}'_t\right) = 0
\end{equation}
and
\begin{equation}
    \sum_{i=1}^{n} \left(\frac{x'_i - p'_{ri}}{\sigma'_i}\right)^2 - 1 = 0,
\end{equation}
where $A$ is a scalar constant. From this, we can define
\begin{equation}
    \varepsilon = \lVert \mathbf{x}' - \mathbf{p}'_r \rVert,
\end{equation}
the uncertainty on $\mathbf{p}'_r$ in the direction of $\mathbf{p}'_t$. Consequently, our final metric, quantifying the retrieval precision, $\mathcal{M}_2$, can be stated as
\begin{equation}
    \mathcal{M}_2 = \frac{\mathcal{M}_1}{\varepsilon} = \frac{\lVert \mathbf{p}'_t - \mathbf{p}'_r \rVert}{\varepsilon}.
    \label{eq:M2}
\end{equation}
Figure \ref{fig:metrics_illustration} shows a diagram illustrating $\mathcal{M}_1$, $\mathcal{M}_2$ and $\varepsilon$ in a two-dimensional parameter space.

\subsection{Informed priors for nested sampler retrieval}
\label{sec:informed_priors_results}
\subsubsection{Classification accuracy}
\label{sec:classification_accuracy}
Since nested samplers draw all their samples directly from the priors, it is crucial for this method that a spectrum is assigned to a class whose parameter distribution ranges include the atmospheric parameters for that spectrum. Nested sampling cannot return accurate posteriors if the spectrum is mis-classified to a class that excludes any of its parameter values. 

For our simulations we define an accurate classification to be one in which the true parameters of the simulated spectrum are contained within the prior volume of the class to which the spectrum is assigned, that is to say that the parameters have a non-zero probability density within the assigned class. Conceivably, noise within real data may result in mis-classification. To investigate this, we generate 5000 simulated spectra with random parameters drawn uniformly from the parameter distributions in Table \ref{tab:parameter values} for each of the spectral resolutions $R=30, 100$, and $300$. We then generate a noisy realisation of the spectrum by applying Gaussian noise within each spectral bin at levels ranging from $0-50$~per~cent of the maximum-to-minimum (``peak-to-trough'') variation in the spectrum. 

\begin{figure*}
    \centering
    \includegraphics[width=\textwidth]{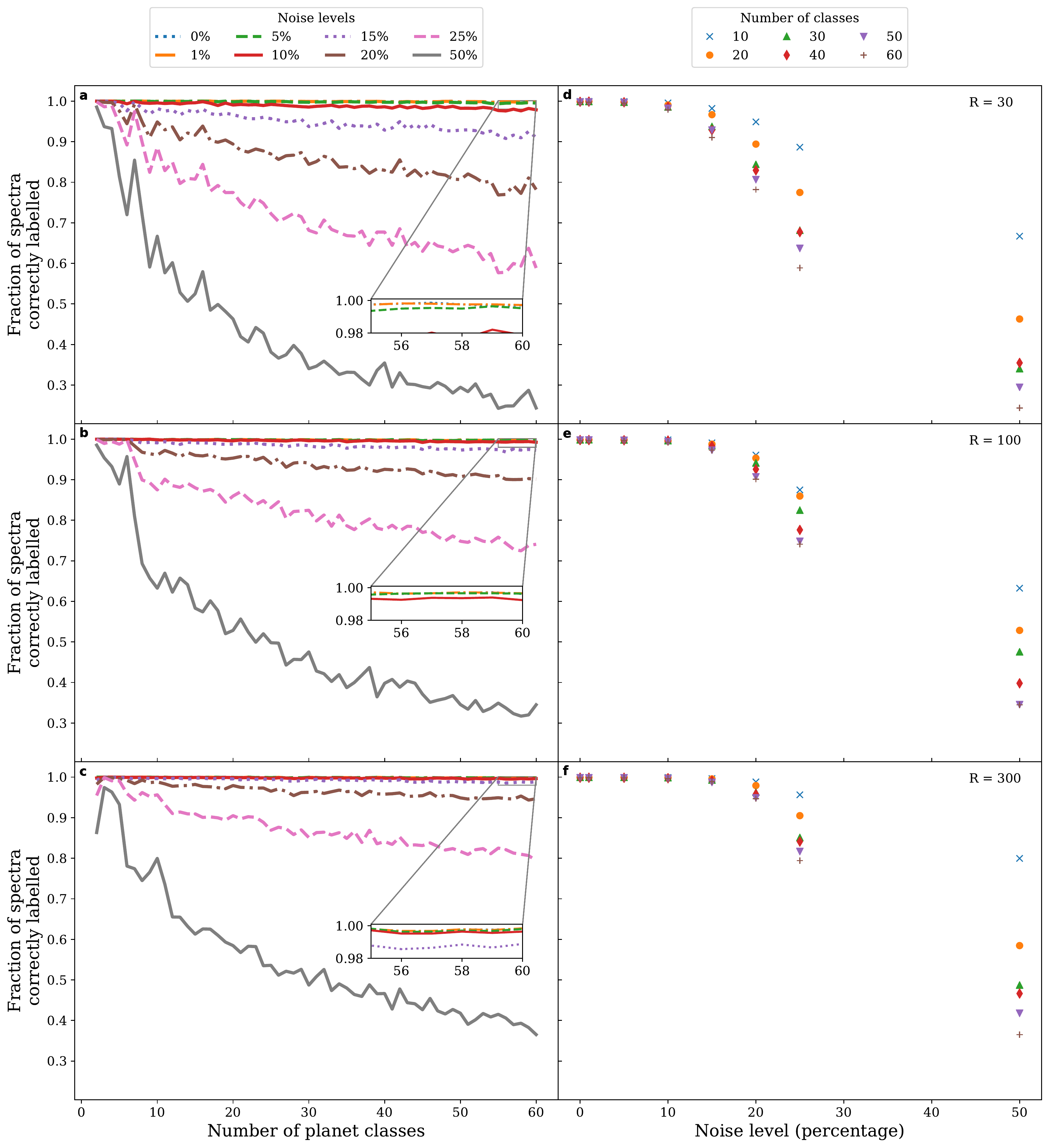}
    \caption{Left (panels \textbf{a, b, c}): The fraction of simulated spectra accurately classified as a function of the number of classes for different noise levels. Results are shown for spectral resolutions of $R=30$ (top), 100 (middle), and 300 (bottom). The noise levels are quoted as a percentage of the dynamic range of the spectra. The zoomed in areas show the differences between the best performing noise levels at the highest number of classes tested. Right (panels \textbf{d, e, f}): The fraction of spectra accurately classified as a function of noise level for classifiers using 10, 20, 30, 40, 50, and 60 classes. These results are based on $5000$ simulated spectra, with parameters drawn uniform randomly from the ranges in Table \ref{tab:parameter values}.}
    \label{fig:classification_accuracy}
\end{figure*}

Figures \ref{fig:classification_accuracy}a-c show the fraction of these 5000 spectra that are correctly classified as a function of the number of classes imposed on the $k$-means clustering classification described in Section~\ref{sec:training_classifier}. Figures \ref{fig:classification_accuracy}d-f show the fraction of the spectra that are correctly labelled as a function of noise level for classifiers with 10, 20, 30, 40, 50, and 60 classes for the same spectral resolutions. 

We can visualise the effect of noise on a spectrum as an uncertainty of position in PCA space. Spectra that lie near a boundary between two classes are more susceptible to being mis-classified. As the number of classes is increased, the number of boundaries also increases, leading to the general negative correlation we see in Figures \ref{fig:classification_accuracy}a-c. However, a spectrum that changes class as a result of noise is only considered mis-classified if the new class assignment does not include all of the physical parameter values associated with the spectrum. Whilst classes are segregated in PCA space, their physical parameter distributions can nonetheless overlap, meaning that the physical parameters of a spectrum can be included within more than one class. Assignment to any of these counts as a valid classification that will enable a successful retrieval.

When increasing the number of classes, $N_\textrm{c}$, we increase the number of boundaries. Roughly speaking, the average volume of a class in PCA space, $\overline{V}_\textrm{c}$, scales as
\begin{equation}
    \overline{V}_\textrm{c} \propto \frac{1}{N_\textrm{c}}.   
    \label{eq:average_prior_volume-classes_relation}
\end{equation}
The fractional change in average volume when adding a class is given by 
\begin{equation}
    \frac{\overline{V}_\textrm{c}}{\overline{V}_\textrm{c}+\Delta \overline{V}_\textrm{c}} = \frac{N_\text{c}}{N_\text{c} + 1},
\end{equation}
and so the average volume of a class is more sensitive to the addition of extra classes at lower $N_\text{c}$. At low $N_\text{c}$ the mis-classification fraction is especially sensitive to the redistribution of class centres in PCA space that results when $N_\textrm{c}$ is varied. This can be seen in the larger fluctuations present for low numbers of classes in Figures~\ref{fig:classification_accuracy}a-c. 

Figures \ref{fig:classification_accuracy}d-f show that there is effectively a cut-off in the noise level where the accuracy of classification rapidly deteriorates. The cut-off point increases with spectral resolution, but the deterioration in accuracy becomes more significant with higher numbers of classes. For $R=30$, the cut-off noise level is around $10$~per~cent, and we suggest therefore that our method is appropriate for spectra with resolution $R\geq30$ and noise less than $10$~per~cent of the peak-to-trough amplitude in the spectrum. With much higher noise levels, the number of classes required to give  comparable accuracy reduces until reaching a limit of one class - the limit where the informed prior spans the whole physical parameter range and is therefore indistinguishable from an uninformed prior. At this point the informed prior no longer offers any practical advantage for retrieval efficiency.

As the spectral resolution increases, the overall accuracy of classification increases. This behaviour is expected, as at higher resolution, there are more spectral features, which lead to more clearly delineated classes in PCA space. At $R=300$, spectra with $15$~per~cent noise have an accurate classification rate over $98$~per~cent, indicating that this method lends itself well to higher resolution spectra, as large numbers of classes can be used on these spectra.

From looking at Equations \ref{eq:iterations-prior_relation} and \ref{eq:average_prior_volume-classes_relation}, we can see that it is desirable to have a large number of different classes. As $\overline{V}_\text{c}$ reduces, the prior volume must similarly reduce, and so with a higher number of classes, a smaller number of iterations will be required by the retrieval, as illustrated by Equation \ref{eq:iterations-prior_relation}. Indeed, if the number of classes is increased greatly, we reach a limit where in the event of no noise a spectrum can be immediately matched to a set of parameters without the need to run any form of retrieval code beyond the classification.

However, in practice real spectra contain noise so some form of retrieval algorithm will always be required. Furthermore, in the presence of noise there is always a non-zero risk of misclassification that can thwart successful retrieval using an informed prior. In practice, the number of classes should be tuned according to the noise level of the data so as to give an acceptable accurate classification rate. Our experience here is that the method is effective for noise levels at or below $10$~per~cent of the spectrum peak-to-trough dynamic range.

The percentage of misclassification is clearly relevant when determining the effectiveness of using informed priors. For misclassified spectra our retrieval will fail, resulting in wasted computational effort as we must repeat the retrieval using a standard uninformed prior. However, when performing retrievals over a large catalogue of objects, a failure rate at the $\sim 1$~per~cent level incurs a time overhead that can still be substantially out-weighed by the time savings of using informed priors for correct classifications. In the next Section we show that our method provides typically a factor two saving in computational efficiency, so a misclassification rate of $1-5$~per~cent does not significantly impact on the advantages of using informed priors.As we show in Section \ref{sec:real_spectra}, classification accuracy can reach $>99.5$~per~cent for JWST-quality spectra. We also discuss and illustrate the fact that the misclassification rate is strongly linked to the degeneracy of surviving models for the given data quality. Hence, in a real-world application, misclassified cases still lead to viable models that cannot be excluded by the data. The misclassification rate should therefore be regarded as a diagnostic of the ability of the data to discriminate between models rather than an indicator that our method necessarily leads to an erroneous solution.

\subsubsection{Effect of informed priors on retrieval}
\label{sec:ret}
To investigate the effect of using informed priors we apply them to transmission spectrum retrieval using nested sampling. We consider three different values for spectral resolution, $R=30, ~100$, and $300$, and two levels of Gaussian noise, corresponding to $1$~per~cent and $10$~per~cent of the spectrum peak-to-trough amplitude. In reality, it is more common to refer to noise levels in parts-per-million (ppm), but in this section we seek to demonstrate the basic effectiveness of our method by using a relative noise scale. We address the application of using our informed priors method on realistic spectra in Section \ref{sec:real_spectra}, when we explore the effectiveness of more limited wavelength ranges, varying spectral resolutions, and higher, absolute noise levels. The $k$-means classifier is set to define 30 classes. We see from Figure \ref{fig:classification_accuracy} that 30 clusters gives $>98$~per~cent accuracy for classification at both noise levels and at all the spectral resolutions under consideration.

\begin{figure*}
    \centering
    \includegraphics[width=\textwidth]{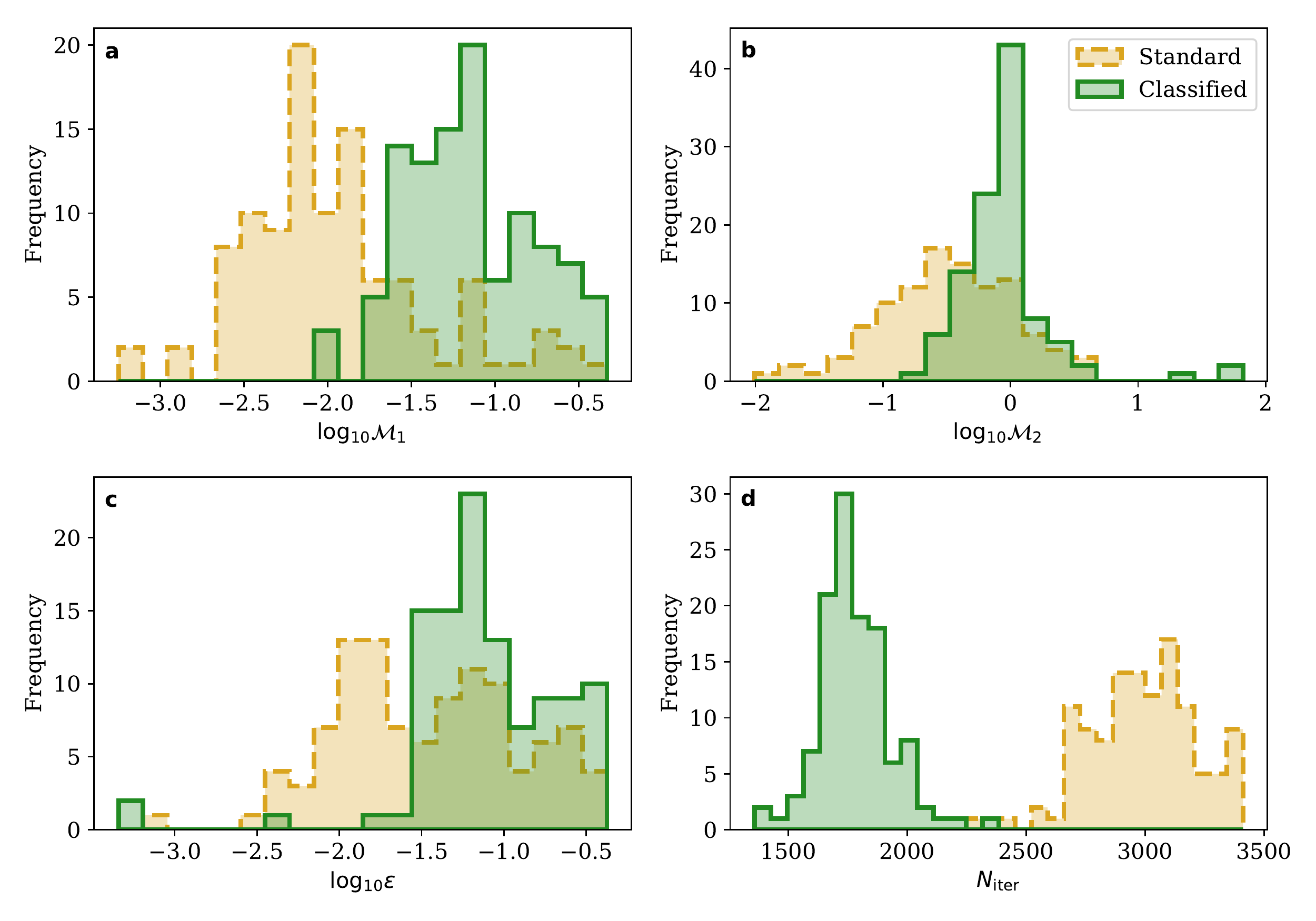}
    \caption{Metric values obtained for retrieval of 120 spectra generated from random parameters with resolution $R=100$ and $1$~per~cent noise. For each of the 120 spectra, nested sampler retrieval was run with and without classifier-informed priors. Dashed yellow histograms denote standard retrieval, whilst solid green histograms denote classified retrieval. There were 30 classes used for classification. \textbf{a} shows the log of the accuracy metric, $\mathcal{M}_1$, whilst \textbf{b} shows the log of the precision metric $\mathcal{M}_2$. \textbf{c} shows the log of the error in the direction of the vector in parameter space linking the true and retrieved parameters. \textbf{d} shows the number of iterations required for the retrieval to reach the stopping criterion shown in Equation~\ref{eq:stopping_criterion}.}
    \label{fig:R100_n0.01_metrics}
\end{figure*}

We generate 120 spectra at each resolution with parameters drawn randomly from the ranges in Table \ref{tab:parameter values}, create a noisy realisation of the model spectrum, and then run two retrievals for each spectrum. One retrieval uses a uniform prior (which we label ``standard'') whilst the other uses the prior obtained from our unsupervised ML classification (which we label ``classified''). Standard retrieval uses the nested sampler retriever packaged with PLATON to obtain the best fit model, along with the metrics  $\mathcal{M}_1$, $\mathcal{M}_2$, $\varepsilon$, and $N_\textrm{iter}$. For classified retrieval we use a slightly edited version of the PLATON code, which is identical except the \texttt{transform\_prior} function is changed to use the parameter distributions extracted from a PCA class as an informed prior, rather than the uninformed standard prior used by the PLATON \texttt{FitInfo} object. We run this classified retrieval and, as for the standard case, obtain values for the metrics. Following \cite{PLATON2019}, the nested sampler we use has 100 live points. Figure \ref{fig:R100_n0.01_metrics} shows the values of these metrics obtained for $R=100$ and $1$~per~cent noise. 

Figure \ref{fig:R100_n0.01_metrics}(d) shows that there is a significant reduction in the number of iterations required for the retrieval to reach its stopping criterion. In the case of $R=100$ and $1$~per~cent noise, there is a $41$~per~cent reduction in the mean number of iterations. The means of the distributions in Figure \ref{fig:R100_n0.01_metrics}(d) are $1779$ and $2991$ for classified and standard retrieval, respectively. This difference of $1212$ iterations corresponds to $121\,200$ fewer likelihood evaluations. On a Macbook Pro with a $2.7$~GHz Intel Core i5 processor, a single run of the PLATON forward model, which must be invoked in any likelihood evaluation, takes $105\pm 9$~ms, based on 1000 calculations. This translates to a reduction in CPU time of around $3.3$~hours.

\begin{figure}
    \centering
    \includegraphics[width=\columnwidth]{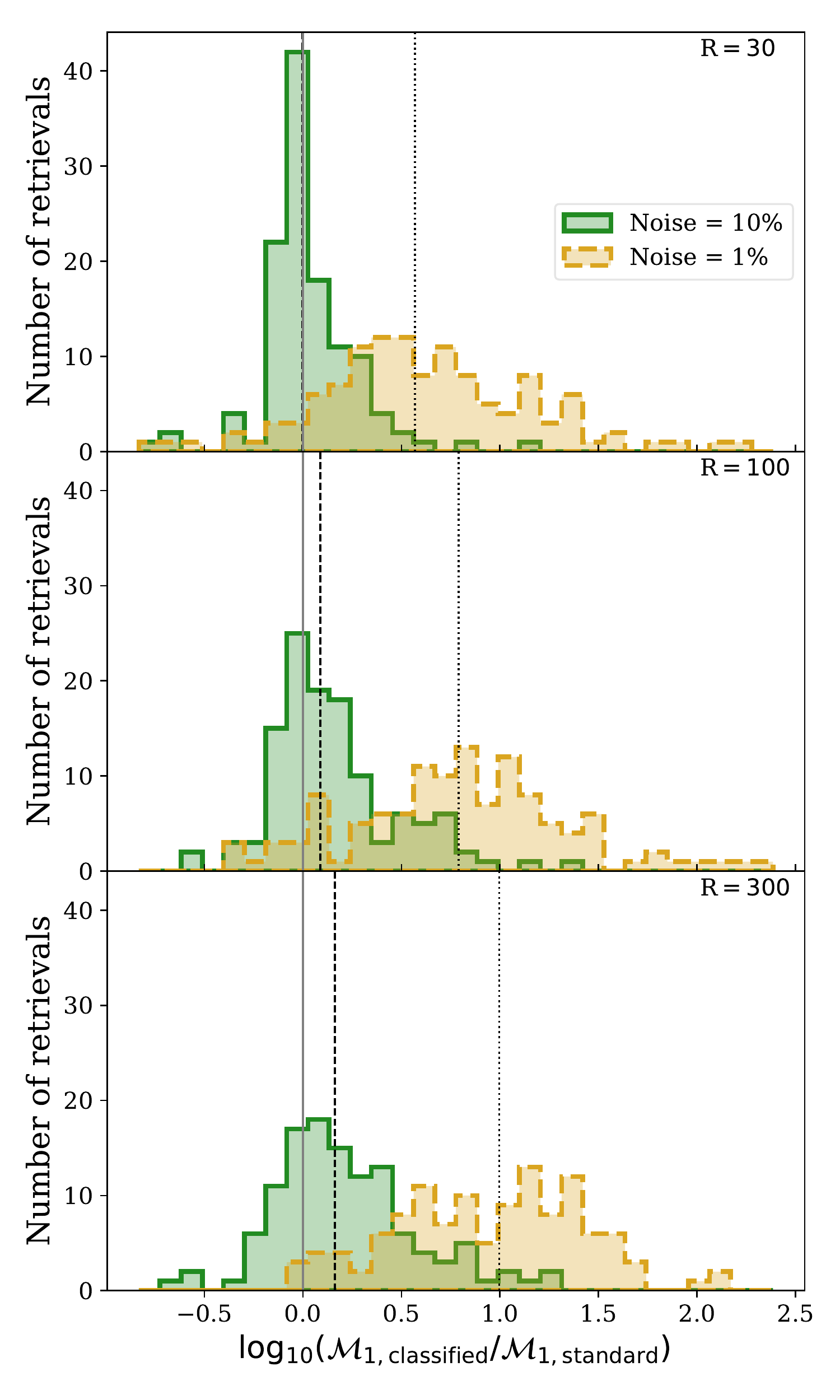}
    \caption{From top to bottom, these plots show the log-ratio of the $\mathcal{M}_1$ values for classified and standard retrieval for $R=30,~100$ and~$300$. The dashed yellow histograms are for retrieval of data with noise at $1$~per~cent and those in solid green show retrieval of data with $10$~per~cent noise. The dotted and dashed vertical line are the median value of the ratio for $1$~per~cent and $10$~per~cent noise, respectively, while the solid vertical line indicates a ratio of $1$.}
    \label{fig:M1_ratios}
\end{figure}

In Figure~\ref{fig:R100_n0.01_metrics}(a) it is noticeable that the standard retrievals report lower $\mathcal{M}_1$ values. $\mathcal{M}_1$ measures the normalised distance between the retrieved and true parameters, something which is ultimately governed by the nested sampling stopping criteria that is based upon the fraction of remaining evidence $\Delta \ln\hat{\mathcal{Z}}$ (Section~\ref{sec:nested_sampling}). Unless the true solution lies close to the peak in the informed prior distribution, where most of the prior mass is contained, we must expect that a stopping criterion based on the fraction of evidence that remains will stop classified retrieval somewhat further from the true solution than for standard retrieval where the prior mass is evenly distributed throughout. Figure \ref{fig:M1_ratios} indicates that this difference in behaviour is more likely to be evident where there is more detail available in the data (higher resolution and/or lower noise). In practice this behavioural difference could be minimised either by allowing for more classes when dealing with higher quality data (so that the determination of how much evidence remains is more locally evaluated) or by modifying the stopping criterion to halt at a fixed absolute value of remaining evidence. In any case, reassuringly, Figure~\ref{fig:R100_n0.01_metrics} confirms that the classified retrieval nonetheless obtains parameters that are typically within the expected distance of the correct solution given the noise level (i.e. centred on $\log_{10} \mathcal{M}_2 \approx 0$). This means that in reality the solutions retrieved by either the standard or classified approach are indistinguishable at the level of the noise within the data. Around 3~per~cent of the classified points have very large $\mathcal{M}_2$ values, which is likely due to the misclassification of the input spectrum. These outliers are the main contributors to the highest $\mathcal{M}_1$ values we see in Figure \ref{fig:R100_n0.01_metrics}a.

\begin{figure}
    \centering
    \includegraphics[width=\columnwidth]{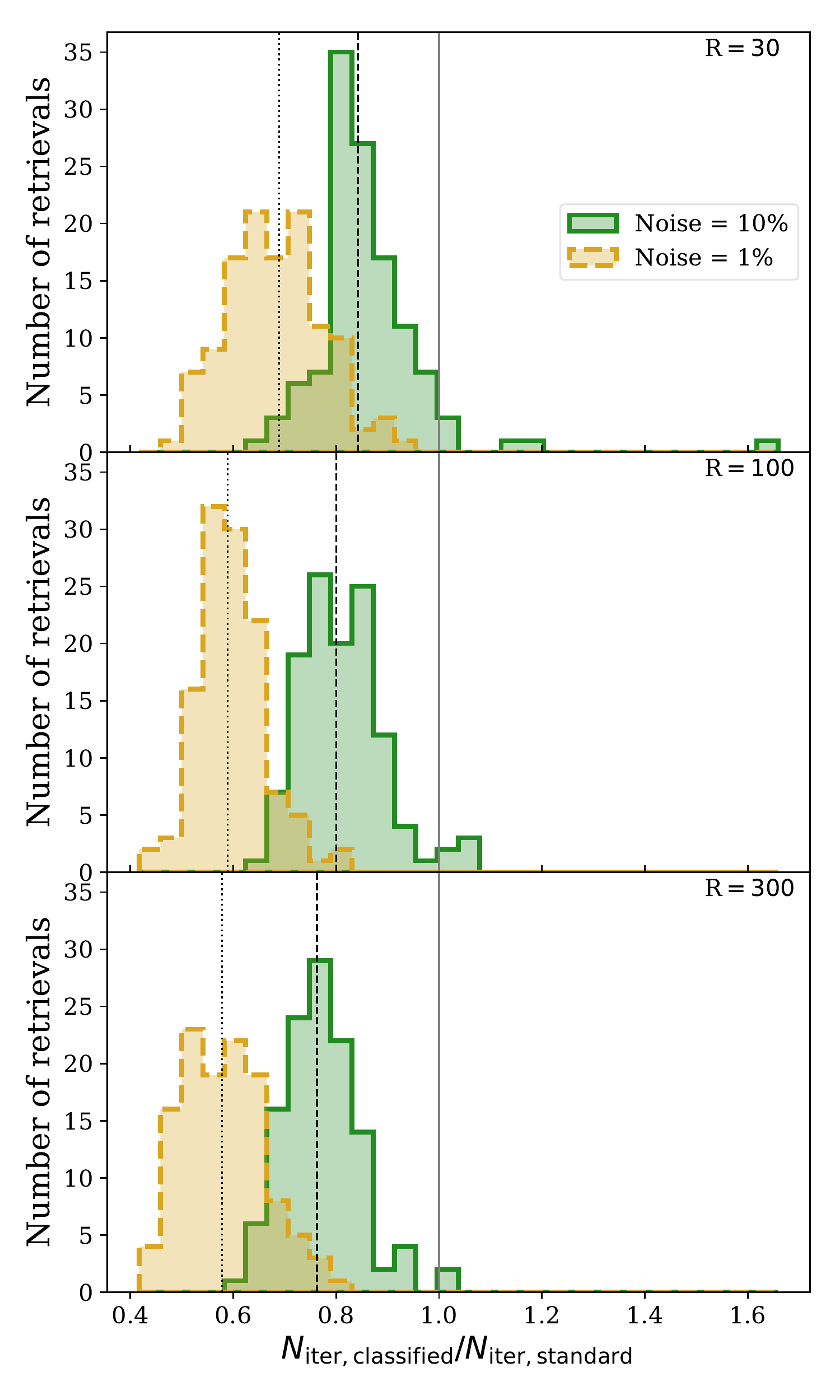}
    \caption{From top to bottom, these plots show ratio of the number of iterations required for retrieval with and without the use of our optimised priors for $R=30$,~$100$,~and~$300$. The dashed yellow histograms are for retrieval of data with noise at $1$~per~cent and those in solid green show retrieval of data with $10$~per~cent noise. The dotted and dashed vertical line are the median value of the ratio for $1$~per~cent noise and $10$~per~cent noise respectively, while the solid vertical line indicates a ratio of $1$.}
    \label{fig:iterations_ratio}
\end{figure}

\begin{figure}
    \centering
    \includegraphics[width=\columnwidth]{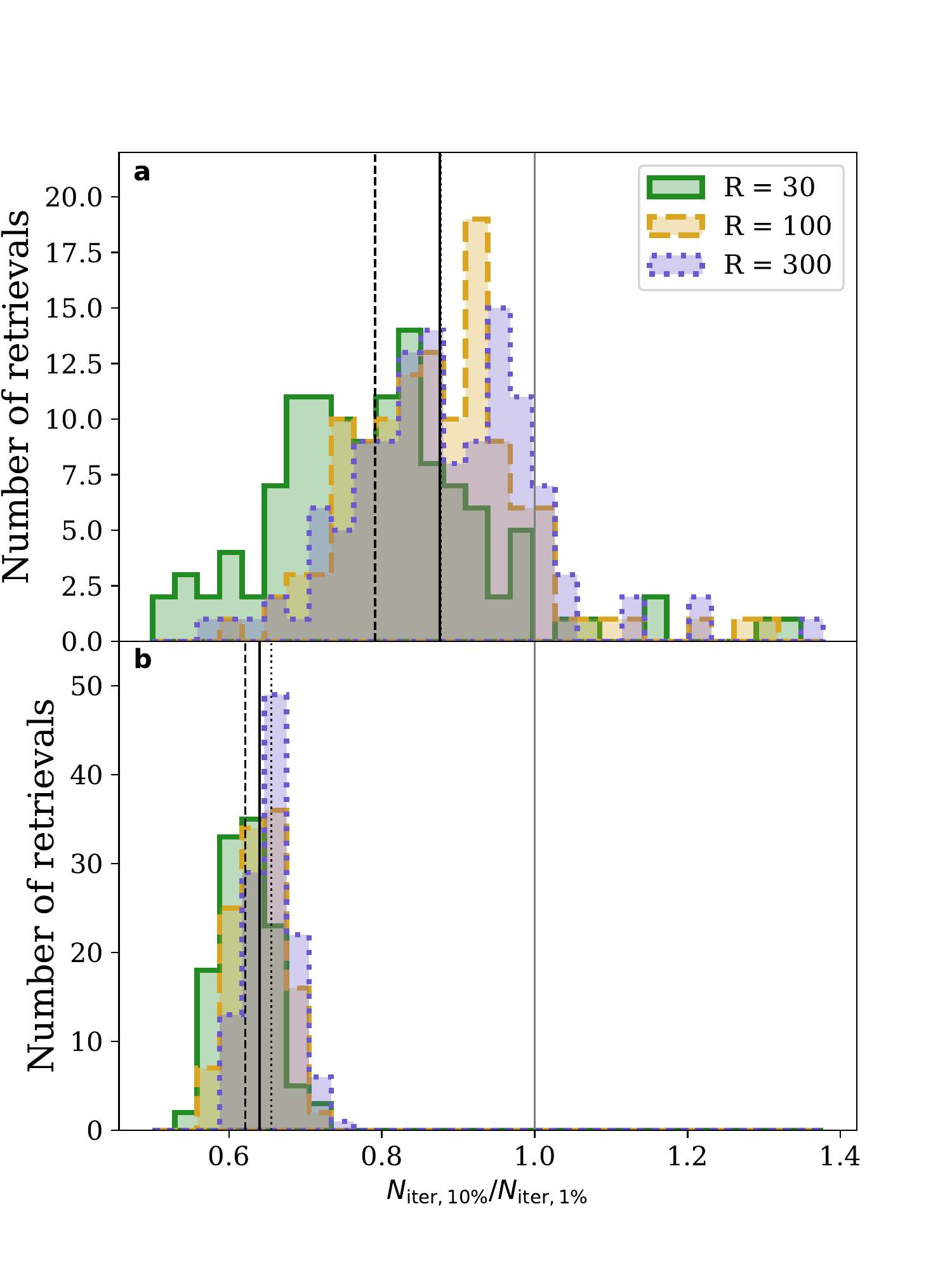}
    \caption{The ratios of the number of iterations required for retrieval with $10$~per~cent and $1$~per~cent noise for (\textbf{a}) classified nested sampler retrieval and (\textbf{b}) standard nested sampler retrieval. The black vertical lines denote the median value of each of the histograms. The solid grey line denotes a ratio of $1$.}
    \label{fig:niter_noise}
\end{figure}

Figure \ref{fig:iterations_ratio} is similar to Figure \ref{fig:M1_ratios}, but instead shows the ratio of $N_\textrm{iter}$ for retrieval with the classified and standard approaches for each of the 6 combinations of resolution and noise. We can see that almost all retrievals are faster with the use of the classifier. There is a small correlation where higher resolution spectra have a larger reduction in iterations, partly due to the effect of the stopping criterion on $\mathcal{M}_1$ discussed above, but also due to  higher resolution spectra containing more structure, reducing degeneracies and hence reducing the number of parameters which must be tested. We also see that noise level has an impact on $N_\textrm{iter}$, with noisier spectra having a ratio closer to one. This is explained through Figure \ref{fig:niter_noise}, which shows the ratio of $N_\textrm{iter}$ for $10$~per~cent and $1$~per~cent noise for each of the resolutions for classified and standard retrieval. From Figure \ref{fig:niter_noise} it is clear that noise rather than resolution is the biggest determinant of $N_\textrm{iter}$. Increasing the noise level results in a reduction of the number of iterations required for retrieval, with a more pronounced effect for standard retrievals. Since $N_\textrm{iter}$ for standard retrievals is reduced by a larger fraction as noise level increases, it is clear that the relative efficiency gain in using a classified approach diminishes as noise increases. However, even for $10$~per~cent noise we note that there is still a median speed-up of $\sim20$~per~cent seen in Figure \ref{fig:iterations_ratio}, meaning that our classified method still out-performs the standard approach. We also see from Figure \ref{fig:M1_ratios} that at higher noise, the $\mathcal{M}_1$ values are more comparable, implying that the accuracy of the two approaches converges as the information content within the data diminishes, as we should expect.

\begin{figure*}
    \centering
    \includegraphics[width=\textwidth]{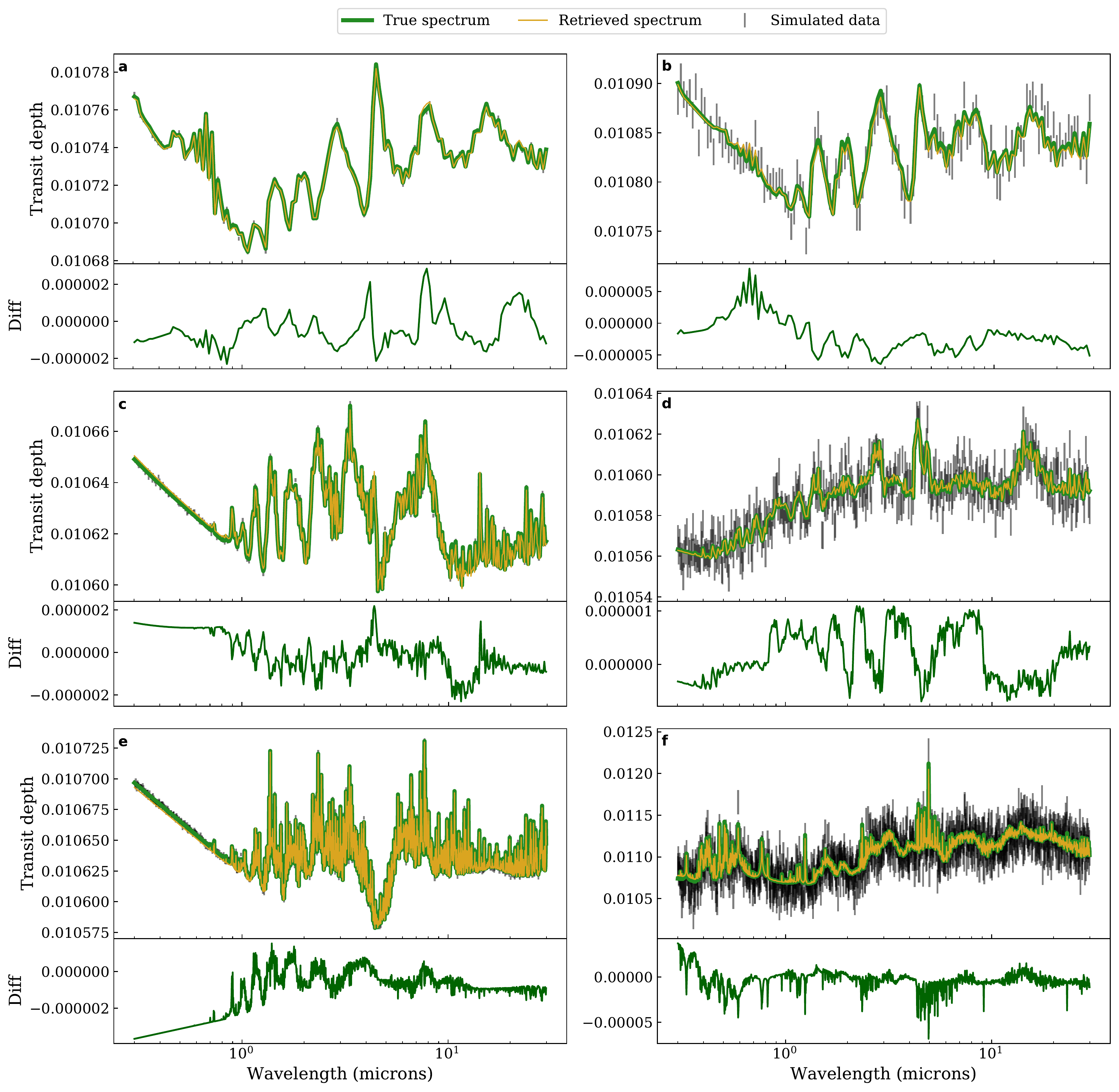}
    \caption{The upper plot of each subplot shows the comparison of a PLATON-generated spectrum with parameters retrieved using classified nested sampling and the true parameters of the simulated spectrum, while the lower shows the difference between the two. Table \ref{tab:spectrum_parameters} shows the true set of parameters for each of the spectra. The data points used for retrieval are shown in black and have had noise added at the level indicated in Table \ref{tab:spectrum_parameters}. The spectra retrieved by the classified approach are shown in yellow. The green spectra are the true spectra, for which we use a thicker line to enable it to be seen under the retrieved spectrum.}
    \label{fig:demo_spectrum}
\end{figure*}

Figure \ref{fig:demo_spectrum} shows some example spectra generated by classified retrieval compared with the true spectrum for a variety of parameters, spectral resolutions and noise levels, the details of which are shown in Table \ref{tab:spectrum_parameters}. We see that the true spectrum and retrieved spectrum are close to indiscernible, with the majority of differences below the $1$~per~cent level. Larger differences are present for very narrow features, but the features are still present in the retrieved spectrum. 

\begin{table*}
    \caption{The true parameters for each of the example spectra in Figure \ref{fig:demo_spectrum}.}
    \label{tab:spectrum_parameters}
    \begin{tabular}{l|c|c|c|c|c|l}
        \hline
         & a & b & c & d & e & f\\
        
        \hline
        Resolution & 30 & 30 & 100 & 100 & 300 & 300 \\
        Noise (per cent) & 0.01 & 0.1 & 0.01 & 0.1 & 0.01 & 0.1\\
        Planet mass ($\mjup$) & $1.68$ & $1.87$ &  $1.59$ & $1.74$ & $0.74$ & $1.78$ \\
        Temperature (K) & $2067$ & $1590$ & $463$ & $709$ & $260$ & $2962$ \\
        Log metallicity & $2.73$ & $1.85$ & $1.66$ & $2.72$ & $-0.51$ & $0.21$ \\
        C/O ratio & $0.22$ & $0.21$ & $1.61$ & $0.54$ & $0.99$ & $1.99$ \\
        Log scattering factor & $3.97$ & $3.81$ & $2.80$ & $-1.88$ & $1.52$ & $-0.78$ \\
        \hline
    \end{tabular}
\end{table*}

\subsection{Application to realistic spectra}
\label{sec:real_spectra}
So far, we have introduced our method for creating informed priors and demonstrated their application to a set of idealised light curves, showing that in the case of low-noise and broad wavelength coverage fewer iterations are required for accurate retrieval. Now we look at the application of this method to spectra of the quality likely to be obtained from current or near-future observatories. The observatories we use are the \textit{Hubble Space Telescope} (\textit{HST}), for observations similar to that of GJ1214b with the WFC3 instrument by \citet{KreidbergGJ1214b}, the planned \textit{Twinkle} mission \citep{Twinkle_mission}, the NIRSpec instrument which will be mounted on the \textit{James Webb Space Telescope} \citep[\textit{JWST},][]{NIRSpec2004}, and the FORS2 instrument mounted on the VLT, as used in the spectroscopic studies by \citet{FORS2_Nikolov}, which uses both the GRIS600B and GRIS600RI grisms available in FORS2. In order to simulate the reality of multi-instrument spectroscopy, we also investigate two different composite spectra by combining observations from 
\begin{enumerate}
    \item\label{HST+Twinkle+JWST}\textit{HST}, \textit{Twinkle}, and NIRSpec, and
    \item\label{HST+FORS2}\textit{HST} and FORS2. 
\end{enumerate}
 The full details on the wavelength ranges, resolutions, and noise levels in ppm for each of these observatories are shown in Table \ref{tab:telescope_spectra_details}. \textit{Twinkle} has three different wavelength bins, which are specified separately.

\begin{table*}
    \caption{The wavelength ranges, spectral resolutions, and noise levels in ppm of the instruments for which spectra were generated and retrieval was run. \textit{Twinkle} has three different wavelength bins, which are specified separately. In addition to these instruments being tested separately, a retrieval test was also run with a composite spectrum of \ref{HST+Twinkle+JWST} \textit{HST}, \textit{Twinkle}, and \textit{JWST}-NIRSpec, and \ref{HST+FORS2} \textit{HST} and FORS2, simulating the use of observations from multiple instruments.}
    \label{tab:telescope_spectra_details}
    \begin{tabular}{l l l l l l l}
        \hline
        Observatory & Low wavelength & High wavelength & Resolution & Noise level \\
        \hline
        \textit{HST} (WFC3) \citep{KreidbergGJ1214b} & $1.1~\mu$m & $1.7~\mu$m & $70$ & 30~ppm \\
        \textit{Twinkle} \citep{Twinkle_instrument} & $0.4~\mu$m & $1~\mu$m & $250$ & 100~ppm\\
        & $1.3~\mu$m & $2.42~\mu$m & $250$ & 100~ppm \\
        & $2.42~\mu$m & $4.5~\mu$m & $60$ & 50~ppm  \\
        \textit{JWST}-NIRSpec \citep{NIRSpec2004} & $0.6~\mu$m & $5.3~\mu$m & $100$ & 30~ppm \\
        FORS2 (GRIS600B and GRIS600RI grisms) \citep{FORS2_Nikolov} & $0.411~\mu$m & $0.81~\mu$m & $60$ & 240~ppm \\ 
        \hline
    \end{tabular}
\end{table*}

For each observatory or observatory combination, we generated $100\,000$ training spectra and 240 test spectra, with atmospheric parameters drawn as before from the ranges in Table \ref{tab:parameter values}. We then tested different numbers of classes within the classifiers in order to optimise the number of classes against classification accuracy, where we define classification accuracy as before in Section \ref{sec:classification_accuracy}. The number of classes used in each classifier and the associated accuracy is shown in Table \ref{tab:observatory_classifier_accuracy}. It is intuitively obvious that the more features which can be present in spectrum, the more distinction a classifier can make between different classes of spectra. Since \textit{Twinkle} and \textit{JWST}-NIRSpec have large wavelength ranges, more spectral features are available to classify, and consequently a high number of classes can be used to obtain an accuracy $>99$~per~cent. Conversely, the wavelength ranges of \textit{HST} and FORS2 are comparatively narrow, which limits the accuracy of a classifier with more classes. With 5 classes, we achieve a classification accuracy of $97.5$~per~cent for \textit{HST}. However, with only 2 classes, the FORS2 classifier can only achieve $82.5$~per~cent accuracy. This is likely due to the significantly higher noise levels present due to FORS2 being a ground-based observatory, as discussed in Section \ref{sec:classification_accuracy}. Keeping 2 classes and combining HST observations with the FORS2 observations leads to a classification accuracy of $92.9$~per~cent, which whilst a substantial improvement, still leads to a misclassification rate of $7$~per~cent. In practice, finding the number of classes required for a given accuracy for a specific instrument can be done by simulating a set of test spectra alongside the training set and trialling classification with varying numbers of classes. This functionality will be made available in {\sc PriorGen} (Hayes et al. in prep), a software package which generalises the use of informed priors in retrieval to applications beyond transmission spectroscopy.

\begin{table}
    \caption{The number of classes, $N_\text{c}$, used in the classifier for each of the tested observatories and observatory combinations, along with the accuracy of the classification of the 240 spectra tested.}
    \label{tab:observatory_classifier_accuracy}
    \begin{tabular}{l l l}
        \hline
        Telescope & $N_\text{c}$ & Classification accuracy \\
        \hline
        \textit{HST} & 5 & $97.5$~per~cent\\
        \textit{Twinkle} & 30 & $99.17$~per~cent\\
        \textit{JWST}-NIRSpec & 30 & $99.58$~per~cent \\
        \textit{HST} + \textit{Twinkle} + NIRSpec & 30 & $99.58$~per~cent \\
        FORS2 & 2 & $82.5$~per~cent\\
        \textit{HST} + FORS2 & 2 & $92.9$~per~cent\\
        \hline
    \end{tabular}
    
\end{table}

\begin{figure}
    \centering
    \includegraphics[width=\columnwidth]{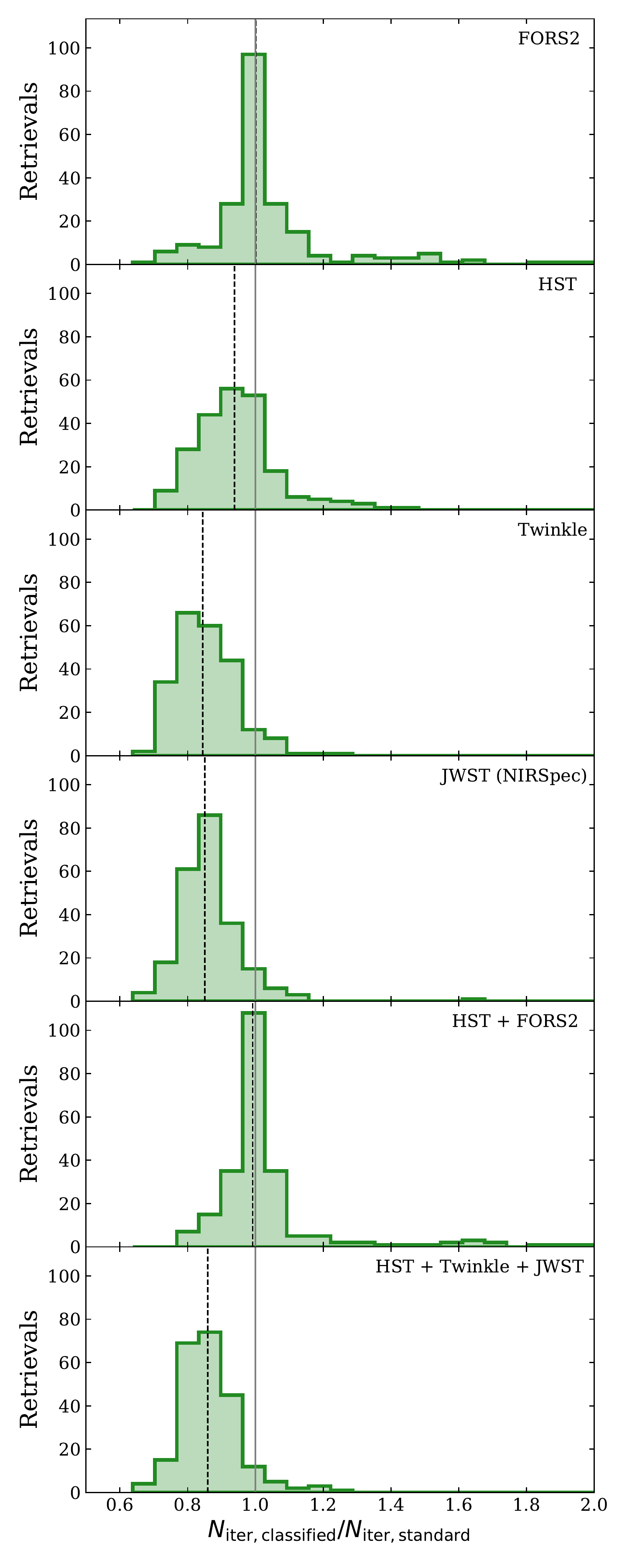}
    \caption{The ratio of the number of observations required to complete retrieval on spectra modelled on different observatories or combinations of observatories with and without the use of our optimised priors. The solid line denotes a ratio of one, while the dashed line shows the median value for each observatory. The classifiers used in the classified retrieval here have the number of classes shown in Table \ref{tab:observatory_classifier_accuracy}.}
    \label{fig:N_iter_telescope}
\end{figure}

\begin{figure}
    \centering
    \includegraphics[width=\columnwidth]{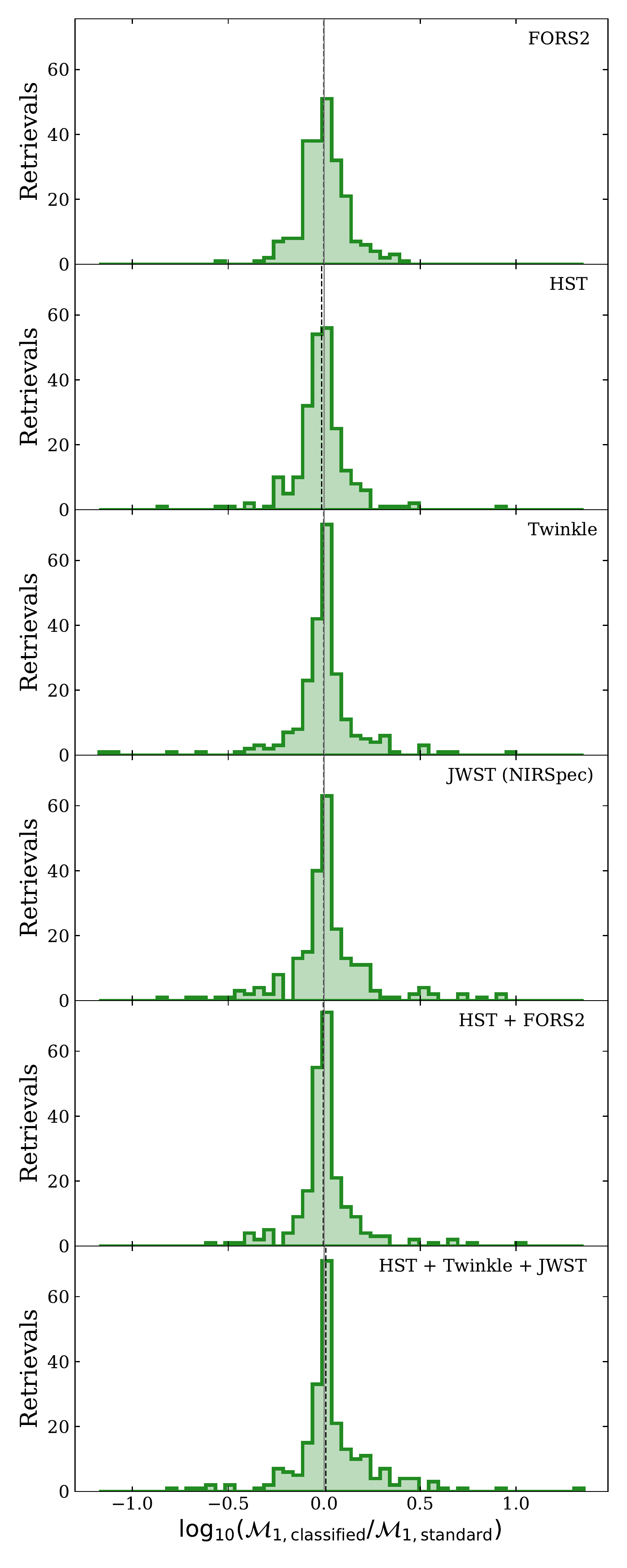}
    \caption{The log-ratio of the $\mathcal{M}_1$ values for each of the classified and standard retrievals of spectra used in Figure \ref{fig:N_iter_telescope}. The solid line denotes a ratio of one, while the dashed lines shows the median value for each observatory.}
    \label{fig:M1_telescope}
\end{figure}

Using classifiers with the number of classes specified in Table \ref{tab:observatory_classifier_accuracy}, we compared the quality of the classified and standard retrieval as before. We should note that we have assumed no systematic baseline offset between observations from different observatories, as we are demonstrating the basic application here. In practical application, this offset term should be included. Figure \ref{fig:N_iter_telescope} compares the number of iterations required for each method for the various observatories. It is clear that, except for the FORS2 cases, our informed priors method still leads to a reduction in the median number of iterations required for a retrieval. In the case of FORS2 and the \textit{HST}-FORS2 combined observations, since we are using only 2 clusters we would expect to see very little average speed-up. We show in Figure \ref{fig:M1_telescope} that the results of the classified and standard retrievals for all the observatories are consistent.

\begin{figure*}
    \centering
    \includegraphics[width=\textwidth]{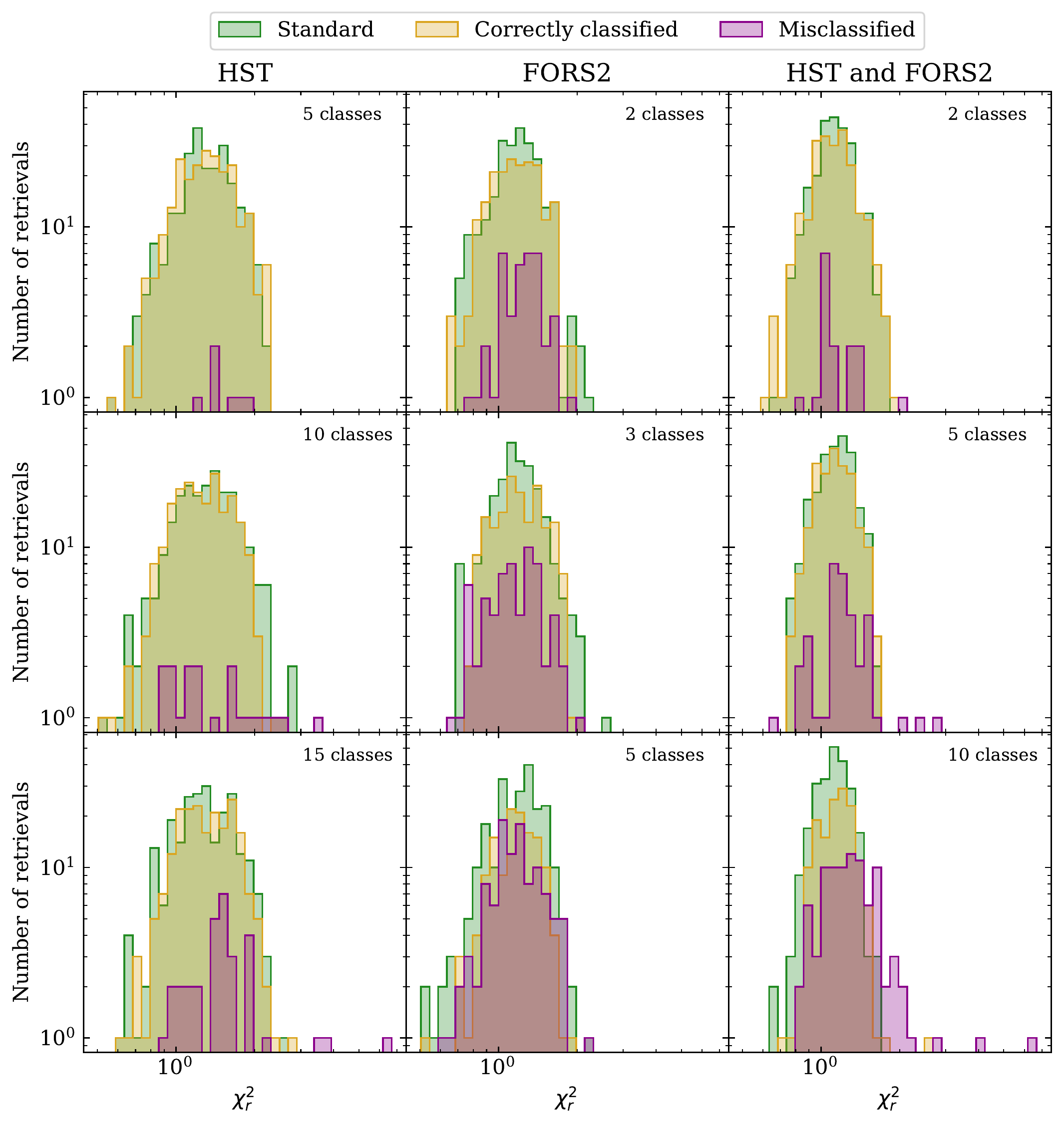}
    \caption{Reduced chi-squared distributions for the retrieval subsets of standard (green), classified with accurate classification (yellow) and classified with a misclassification (purple) for spectra simulated for \textit{HST}, FORS2, and the combination of the two. Accurate classification is defined by the known true parameters of a spectrum being contained within the parameter distribution of the class to which the tested spectrum is assigned. We show results for different numbers of classes to demonstrate the improved accuracy of the classifier with fewer classes, but also to show that even in the case where many spectra are misclassified, viable models with $\chi^2_r < 2$ are still recovered.}
    \label{fig:chi2}
\end{figure*}

\begin{figure*}
    \centering
    \includegraphics[width=\textwidth]{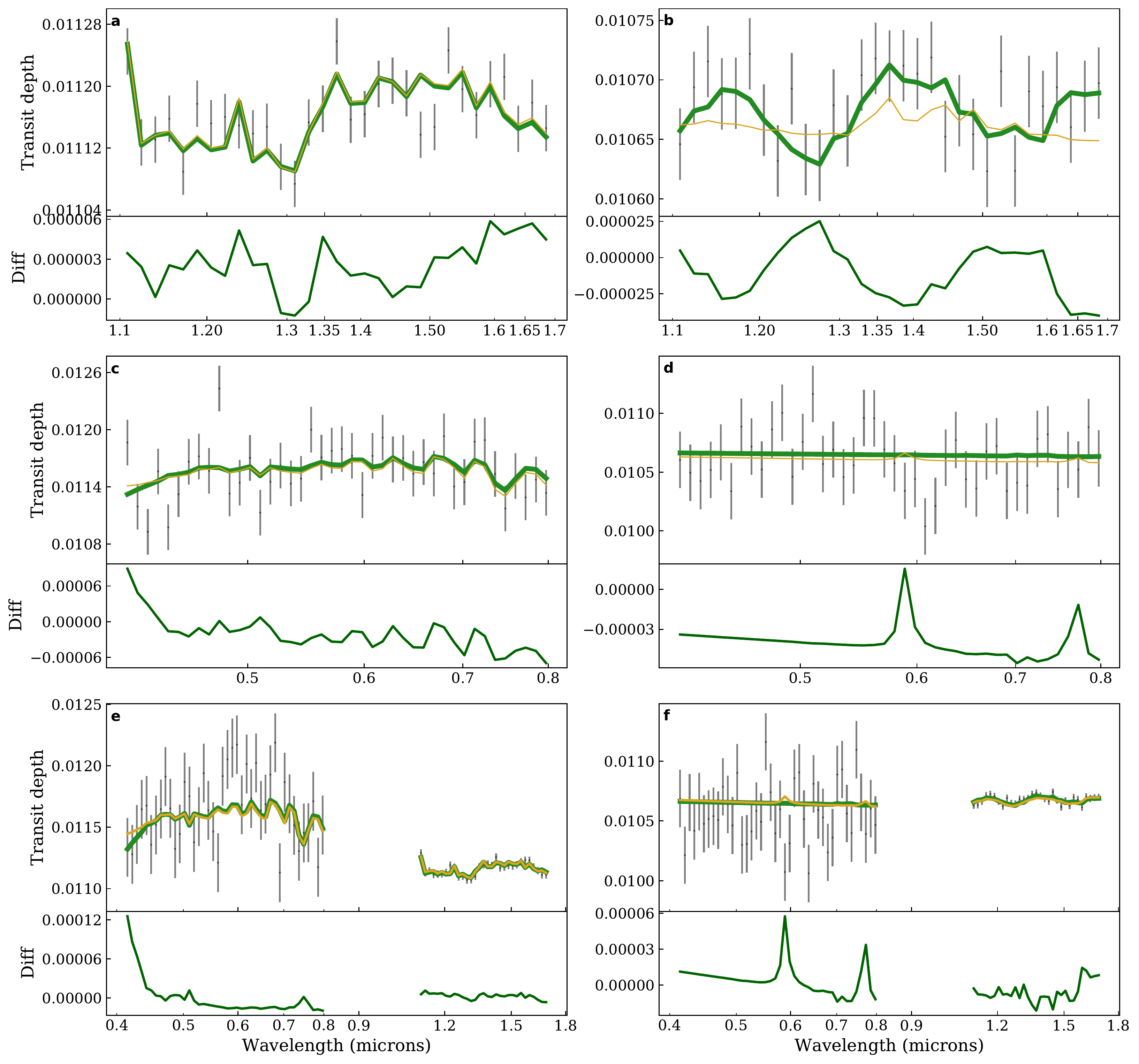}
    \caption{As in Figure \ref{fig:demo_spectrum}, each of these plots shows a comparison of the best-fit model retrieved using our classified nested sampling (in yellow) and a spectrum generated with the true atmospheric parameters (in green). The data points shown are the true spectrum simulated with Gaussian noise at $30$~ppm for \textit{HST} and $240$~ppm for FORS2. Each row shows the spectra simulated for a different observatory: (\textbf{a}, \textbf{b}) show \textit{HST},  (\textbf{c}, \textbf{d}) show FORS2, and (\textbf{e}, \textbf{f}) show a combination of \textit{HST} and FORS2. The spectra on the left (\textbf{a}, \textbf{c}, \textbf{e}) share the same true atmospheric parameters (Planet mass $0.649~\mjup$, temperature $2454$~K, log metallicity $2.55$, C/O ratio $0.23$, log scattering factor $1.65$), and were all accurately classified through our machine learning method. The spectra on the right (\textbf{b}, \textbf{d}, \textbf{f}) also share atmospheric parameters (Planet mass $0.717~\mjup$, temperature $410$~K, log metallicity $2.01$, C/O ratio $0.61$, log scattering factor $2.51$) and were all misclassified by the machine learning algorithm.}
    \label{fig:demo_telescope_spectra}
\end{figure*}

For each of these observatories, we have investigated the distributions of the reduced chi-squared values ($\chi^2_r$) obtained within each retrieval, for the three subsets of standard, classified with accurate classification, and classified with misclassification. The original motivation for this was to investigate if there was a particular value of $\chi^2_r$ that could be used as a diagnosis that a spectrum could have been badly classified, and as such flag it for a potential re-run with more open priors. As such, we focus here on \textit{HST} and FORS2, since Twinkle and \textit{JWST}-NIRSpec have very high classification accuracy. For each telescope, and the combination of the two, we ran retrieval on the same 240 spectra from before, with different numbers of classes within the classifier. Figure \ref{fig:chi2} shows the resulting $\chi^2_r$ distributions for each of these runs, for a varying number of classes. We see that standard and correctly classified runs generally have $\chi^2_r < 2$, which is indicative of a well-fitting model. However, we also find that misclassified runs also tend to have $\chi^2_r < 2$, and so we conclude that, in the real world where a `truth' value is unknown, these models are perfectly acceptable fits. There are a few instances where misclassified models show a poor fit to the data ($\chi^2_r > 2$), which is when we would recommend the re-running of retrieval with wider priors. However, it is clear that in the majority of instances where the classification has failed to correctly identify a prior space which contains the known true values of spectral parameters, the chosen prior space still contains an acceptable model. To illustrate this, we show some example best-fit spectra from correctly classified and misclassified spectra in Figure \ref{fig:demo_telescope_spectra}. We can therefore conclude that the use of our method is limited only by the quality of the data, as is the case for any retrieval algorithm.

\section{Summary and conclusions}
\label{sec:conclusions}
As the number of exoplanet targets suitable for atmospheric follow-up studies is set to increase rapidly over the next few years, we are required to become more selective in target choice and more efficient in our analysis of the data as we move from an era of being target-starved to one in which we will be starved of observing resources for follow-up.

The \textit{Spectroscopy and Phototmetry of Exoplanet Atmospheres Research Network} (SPEARNET) is using an automated approach to target selection and is employing a worldwide telescope network to follow up selected targets. We are also working on methods that can improve atmosphere model retrieval and which can facilitate using retrieval information as part of the criteria for selecting further observations. 

In the present study we have used an unsupervised machine-learning (ML) classifier to construct informed priors suitable for multi-dimensional model retrieval methods like nested sampling, which are used in atmospheric retrieval codes such as PLATON \cite{PLATON2019}. Construction of an informed prior can provide significant retrieval efficiency gains over the standard uninformed prior when fitting a trusted model repeatedly to a large catalogue of objects. Quite apart from exoplanet studies, this is a situation that is being increasingly encountered in astrophysics, and in physics more generally, in the era of ``big data''.

Our approach to constructing an informed prior suitable for retrieval of exoplanet atmosphere models includes the following steps:
\begin{enumerate}
    \item We used PLATON to generate a grid of forward model spectra spanning the relevant range of exoplanet physical parameter space. 
    \item We used a $k$-means clustering algorithm for unsupervised classification of the models.
    \item In order to enable the $k$-means classifier to work efficiently we performed a principal component analysis (PCA) to compress the spectral information down from 461 wavelength bins to just 10 PCA components, with minimal loss of information. Classification was then conducted in PCA space.
    \item We generated test spectra using PLATON, first demonstrating the basic application of the method to spectra covering the full wavelength range of PLATON, convolved to a range of spectral resolutions with added noise, and then to spectra designed to simulate observations from a range of observatories.. The spectra were then passed through the trained classifier to identify which PCA class they belonged to. This process is almost instantaneous and we found it to be highly reliable ($>98$~per~cent) for resolutions $R = 30-300$ and noise levels up to $10$~per~cent of the peak-to-trough amplitude of the spectrum. The process was found to be even more reliable ($>99$~per~cent) for simulated spectra for the upcoming Twinkle and JWST missions.
    \item We are able to extract the parameter distributions corresponding to the PCA class of the test spectrum and use them as an informed prior for nested sampling retrieval of the physical parameters for the spectrum. We showed that we typically improve retrieval efficiency by a factor of two in doing so, with minimal loss of retrieval accuracy, as judged by the difference between retrieved parameters and those used to generate the test spectrum. 
\end{enumerate}
Data noise, rather than spectral resolution, is the main determinant in the success of this method. Whilst some test spectra were incorrectly classified, we found that in these cases a viable model with $\chi^2_r < 2$ is generally still recovered. Consequently only best-fit models with $\chi^2_r > 2$ would require a re-run of retrieval using a standard uniform prior. Since this is $< 1$~per~cent of tested spectra, our method still saves substantial time overall.

SPEARNET is currently in the process of writing a Python 3 package called  {\sc PriorGen} (Hayes et al. in prep) which generalises this method for use in nested sampling retrieval to applications beyond transmission spectroscopy.

Lastly, when moving from a uniform prior to an informed prior approach, it is also possible to embed additional information within the prior that may come from other observations. For example, measurements or constraints on planet mass or equilibrium temperature can easily be folded in by using the measurement and associated error to construct an observational prior that can be incorporated into the existing prior generated from classification. This can either provide further efficiency gains in the retrieval (due to the reduction of the prior mass) or, potentially more usefully, a consistency check with the other measurements by comparing the similarity of retrieval solutions with and without the embedded observational prior.

\section*{Acknowledgements}
The authors would like to thank the anonymous reviewer for their insight and guidance.

JH would like to thank Dr. Alex Clarke and Dr. Therese Cantwell for their assistance and expertise in initial discussions of this project. 

JH and JSM are supported by PhD studentships from the United Kingdom's Science and Technology Facilities Council (STFC). EK and IM are supported by the STFC grant  ST/P000649/1. This work is also supported by a National Astronomical Research Institute of Thailand (NARIT) research grant.




\bibliographystyle{mnras}
\bibliography{informed_priors_paper}




\bsp	
\label{lastpage}
\end{document}